\documentclass[aps,prb,reprint,superscriptaddress]{revtex4-1}

\usepackage{graphicx}% Include figure files
\usepackage{dcolumn}% Align table columns on decimal point
\usepackage{bm}% bold math

\begin{document}

%\preprint{}

\title{Fermi surface in KFe$_2$As$_2$ determined via de Haas-van Alphen oscillation measurements}


%\author{}
%\thanks{}
%\affiliation{}
\author{Taichi Terashima}
\affiliation{National Institute for Materials Science, Tsukuba, Ibaraki 305-0003, Japan}
\affiliation{JST, Transformative Research-Project on Iron Pnictides (TRIP), Chiyoda, Tokyo 102-0075, Japan}
\author{Nobuyuki Kurita}
\altaffiliation[Present address: ]{Department of Physics, Tokyo Institute of Technology, Meguro, Tokyo 152-8551, Japan}
\affiliation{National Institute for Materials Science, Tsukuba, Ibaraki 305-0003, Japan}
\affiliation{JST, Transformative Research-Project on Iron Pnictides (TRIP), Chiyoda, Tokyo 102-0075, Japan}
\author{Motoi Kimata}
\affiliation{Institute for Solid State Physics, University of Tokyo, Kashiwa, Chiba 277-8581, Japan}
\affiliation{JST, Transformative Research-Project on Iron Pnictides (TRIP), Chiyoda, Tokyo 102-0075, Japan}
\author{Megumi Tomita}
%\author{Kota Kodama}
\author{Satoshi Tsuchiya}
%\author{Hidetaka Satsukawa}
%\author{Atsushi Harada}
%\author{Kaori Hazama}
\altaffiliation[Present address: ]{Institute for Solid State Physics, University of Tokyo, Kashiwa, Chiba 277-8581, Japan}
\affiliation{National Institute for Materials Science, Tsukuba, Ibaraki 305-0003, Japan}
\author{Motoharu Imai}
\affiliation{National Institute for Materials Science, Tsukuba, Ibaraki 305-0003, Japan}
\affiliation{JST, Transformative Research-Project on Iron Pnictides (TRIP), Chiyoda, Tokyo 102-0075, Japan}
\author{Akira Sato}
\affiliation{National Institute for Materials Science, Tsukuba, Ibaraki 305-0003, Japan}
\author{Kunihiro Kihou}
\author{Chul-Ho Lee}
\author{Hijiri Kito}
\author{Hiroshi Eisaki}
\author{Akira Iyo}
\affiliation{JST, Transformative Research-Project on Iron Pnictides (TRIP), Chiyoda, Tokyo 102-0075, Japan}
\affiliation{National Institute of Advanced Industrial Science and Technology (AIST), Tsukuba, Ibaraki 305-8568, Japan}
\author{Taku Saito}
\affiliation{Department of Physics, Chiba University, Chiba 263-8522, Japan}
\author{Hideto Fukazawa}
\author{Yoh Kohori}
\affiliation{JST, Transformative Research-Project on Iron Pnictides (TRIP), Chiyoda, Tokyo 102-0075, Japan}
\affiliation{Department of Physics, Chiba University, Chiba 263-8522, Japan}
\author{Hisatomo Harima}
\affiliation{JST, Transformative Research-Project on Iron Pnictides (TRIP), Chiyoda, Tokyo 102-0075, Japan}
\affiliation{Department of Physics, Graduate School of Science, Kobe University, Kobe, Hyogo 657-8501, Japan}
\author{Shinya Uji}
\affiliation{National Institute for Materials Science, Tsukuba, Ibaraki 305-0003, Japan}
\affiliation{JST, Transformative Research-Project on Iron Pnictides (TRIP), Chiyoda, Tokyo 102-0075, Japan}


\date{\today}

\begin{abstract}
We have completely determined the Fermi surface in KFe$_2$As$_2$ via de Haas-van Alphen (dHvA) measurements.
Fundamental frequencies $\epsilon$, $\alpha$, $\zeta$, and $\beta$ are observed in KFe$_2$As$_2$.
The first one is attributed to a hole cylinder near the X point of the Brillouin zone, while the others to hole cylinders at the $\Gamma$ point.
We also observe magnetic breakdown frequencies between $\alpha$ and $\zeta$ and suggest a plausible explanation for them.
The experimental frequencies show deviations from frequencies predicted by band structure calculations.
Large effective masses up to 19 $m_e$ for $B \parallel c$ have been found, $m_e$ being the free electron mass.
The carrier number and Sommerfeld coefficient of the specific heat are estimated to be 1.01 -- 1.03 holes per formula unit and 82 -- 94 mJmol$^{-1}$K$^{-2}$, respectively, which are consistent with the chemical stoichiometry and a direct measure of 93 mJmol$^{-1}$K$^{-2}$ [H. Fukazawa \textit{et al}., J. Phys. Soc. Jpn. \textbf{80SA}, SA118 (2011)].
The Sommerfeld coefficient is about 9 times enhanced over a band value, suggesting the importance of low-energy spin and/or orbital fluctuations, and places KFe$_2$As$_2$ among strongly correlated metals.
We have also performed dHvA measurements on Ba$_{0.07}$K$_{0.93}$Fe$_2$As$_2$ and have observed the $\alpha$ and $\beta$ frequencies.
\end{abstract}

%\keywords{}

\maketitle


\newcommand{\ud}{\mathrm{d}}

\section{Introduction}
Since the discovery of superconductivity at $T_c$ = 26 K in LaFeAs(O, F) by Kamihara \textit{et al}.,\cite{Kamihara08JACS} iron-pnicitde/selenide high-$T_c$ superconductivity has been a center of activity in the condensed matter physics community.\cite{[{For reviews, see }] Ishida09JPSJ_review, *Johnston10AdvPhys}
Many different materials have been synthesized and $T_c$ has quickly been raised up to $T_c \sim$ 55 K.\cite{Kito08JPSJ, Ren08CPL, Yang08SST, Wang08EPL}
The superconducting pairing mechanism is still under debate;\cite{Hirschfeld11RPP} spin fluctuations on the one hand and orbital fluctuations on the other hand.
Basically, the former approach predicts an $s_{\pm}$ gap that changes sign between electron and hole Fermi surface (FS) pockets,\cite{Mazin08PRL, Kuroki08PRL} while the latter predicts an $s_{++}$ gap without sign change.\cite{Kontani10PRL} 

Ba$_{1-x}$K$_x$Fe$_2$As$_2$ (Ref.~\onlinecite{Rotter08PRL}) is one of the most studied systems.
High-quality single crystals can be grown by flux method.
The parent material BaFe$_2$As$_2$ orders antiferromagnetically below the N\'eel temperature $T_N$ = 140 K (Ref.~\onlinecite{Rotter08PRB}) and is a moderately correlated metal with mass enhancements $m^*/m_{band}$ of 2 -- 3 in the ground state, $m^*$ and $m_{band}$ being the effective and the band mass, respectively.\cite{Analytis09PRB, Terashima11PRL}
As K is substituted for Ba, the antiferromagnetism is suppressed and disappears at $x \sim 0.2$, and superconductivity appears.\cite{Rotter08ACIE}
$T_c$ reaches 38 K at optimal doping $x \sim 0.4$.\cite{Rotter08PRL}
$T_c$ does not disappear till $x$ = 1;\cite{Sasmal08PRL, Rotter08ACIE, Chen09EPL} KFe$_2$As$_2$ is a superconductor with $T_c$ = 3.4 K.\cite{Kihou10JPSJ}

Intriguingly, the superconducting gap structure seems to change with $x$ in Ba$_{1-x}$K$_x$Fe$_2$As$_2$.
Near the optimal doping, a fully gapped $s$-wave superconductivity has been indicated by penetration depth,\cite{Hashimoto09PRL} specific heat,\cite{Mu09PRB, Popovich10PRL} thermal conductivity,\cite{Luo09PRB} and NMR measurements.\cite{Yashima09JPSJ} 
On the other hand, a nodal gap structure in KFe$_2$As$_2$ has been suggested by NMR, specific heat,\cite{Fukazawa09JPSJ_KFA} penetration depth\cite{Hashimoto10PRB} and thermal conductivity measurements.\cite{[] [{.  However, note Ref.~\onlinecite{Terashima10PRL_comment}.}] Dong10PRL, Terashima10PRL_comment, Reid12PRL}
A laser angle-resolved photoemission spectroscopy (laser ARPES) and an NMR study as a function of the composition $x$ have found a drastic change in the gap structure around $x$ = 0.6.\cite{Malaeb12PRB, Hirano12JPSJ}
Although it has been under debate whether the superconducting state is an $s$-wave or $d$-wave one,\cite{Reid12PRL} the octet-line nodes of the gap recently observed in another laser ARPES study\cite{Okazaki12Science} are compatible with an $s$-wave state with accidental nodes.
Theoretical studies based on the spin-fluctuations approach suggest a variety of gap structures with $s$- or $d$-wave symmetries depending on band and interaction parameters.\cite{Thomale11PRL, Maiti11PRL, Suzuki11PRB}

In this context, precise determination of the electronic structure near the Fermi level $E_F$ in KFe$_2$As$_2$ is highly desirable.
Previously, we reported de Haas-van Alphen (dHvA) measurements on KFe$_2$As$_2$ in a letter.\cite{Terashima10JPSJ}
We observed one small FS cylinder and two relatively large ones.
With the aid of band structure calculations, they were assigned to a hole cylinder near the X point of the Brillouin zone (BZ) and hole cylinders at the $\Gamma$ point, respectively.
We thought that the largest hole cylinder at $\Gamma$ expected from band structure calculations was not observed.
Recently, we have observed dHvA oscillations in Ba$_{0.07}$K$_{0.93}$Fe$_2$As$_2$.
Interestingly, the largest FS cylinder is clearly observed.
Motivated by this observation, we have reexamined dHvA oscillations in KFe$_2$As$_2$, and have noticed that the largest cylinder is also observed in KFe$_2$As$_2$.
Thus we have completely determined the Fermi surface in KFe$_2$As$_2$ via dHvA measurements.
The mass enhancement amounts to 9, which is much larger than the overall band-width renormalization of 2 estimated in an early ARPES study\cite{Sato09PRL} and indicates the importance of low-energy spin and/or orbital fluctuations.
In addition, we explain magnetic breakdown frequencies observed between the $\alpha$ and $\zeta$ orbits assuming eight magnetic breakdown junctions, the positions of which resemble those of the octet-line nodes.
%In this full paper, we give a detailed account of the dHvA measurements on KFe$_2$As$_2$ and Ba$_{0.07}$K$_{0.93}$Fe$_2$As$_2$.

\section{Experiments, Lifshitz-Kosevich formula, and Yamaji model}
High-quality single crystals of KFe$_2$As$_2$ and Ba$_{0.07}$K$_{0.93}$Fe$_2$As$_2$ were grown by a self flux method as described in Ref.~\onlinecite{Kihou10JPSJ}.
Large residual resistivity ratios of more than 450 and of $\sim$70 were observed for KFe$_2$As$_2$ and K$_{0.93}$Ba$_{0.07}$Fe$_2$As$_2$, respectively.\cite{Kihou10JPSJ}
In the case of KFe$_2$As$_2$, nine different crystals were measured and gave consistent results.
We describe results obtained for the most thoroughly measured sample.

The dHvA measurements were performed in a dilution refrigerator and superconducting magnet by using the field modulation technique.\cite{Shoenberg84}
The modulation frequency and amplitude were mostly $f$ = 67.1 Hz and $b$ = 10.4 mT, respectively, and the detection was made at the second harmonic (2$f$).
A sample was placed in a balanced pick-up coil with its $c$ axis parallel to the coil axis.
The field direction measured from the $c$ axis is denoted by $\theta$, and if necessary a subscript is attached to indicate the field-rotation plane.
The same setup was used to measure ac magnetic susceptibility.

The dHvA magnetization oscillation $M_{osc}$ due to an extremal cyclotron orbit normal to $B$ enclosing the $k$-space area $A$ is given by\cite{Shoenberg84}
\begin{equation}
M_{osc} =  \sum_{r=1}^{\infty} a_r \sin\left(\frac{2\pi rF}{B}+\phi_r\right),
\end{equation}
where
\begin{eqnarray}
a_r \propto \frac{FB^{1/2}}{\mu^*|A''|^{1/2}}r^{-3/2} R_{T, r} R_{D, r} R_{s, r},
\\
R_{T, r}= \frac{rK\mu^* T/B}{\sinh(rK\mu^* T/B)},
\\
R_{D, r}=\exp(-rK\mu^* x_D^*/B),
\\
R_{s, r}=\cos(r\pi S).
\end{eqnarray}
Here $\mu^*= m^*/m_e$, $m_e$ being the free electron mass.
The effective mass $m^*$ is enhanced over the band mass $m_{band}$ by electron-phonon and electron-electron interactions.
The frequency $F$ is given by $F = (\hbar/2\pi e)A$, and $\phi_r$ is the phase.
Not only the fundamental frequency $F$ but also its harmonics ($r > 1$) appear in $M_{osc}$.
$|A''|$ is the curvature factor: $A'' = \partial^2 A/ \partial \kappa^2 $, where $\kappa$ is the wave number along $B$.
$R_{T, r}$ is the temperature reduction factor, where $K$ is a constant (14.69 T/K).
The Dingle factor $R_{D, r}$ describes the influence of disorder/impurity scattering, $x_D^*$ being the Dingle temperature.
We can determine $m^*$ and $x_D^*$ by fitting $R_{T, r}$ and $R_{D, r}$ to $T$- and $B$-dependences of experimental oscillation amplitudes at constant $B$ and $T$, respectively.
The spin reduction factor $R_{s, r}$ is due to the interference between oscillations from up- and down-spin electrons, and the spin-splitting parameter $S$ may be expressed as $S= (1/2)g_{eff}\mu^*$ with an effective $g$ factor $g_{eff}$.

With the present setup and second-harmonic detection, amplitudes of oscillations in the detected voltage, $v_r$, are related to $a_r$ as
\begin{equation}
v_r \propto (\cos \theta - \frac{1}{F}\frac{\ud F}{\ud \theta} \sin \theta)J_2 (r\lambda) a_r,
\end{equation}
where $J_2$ is the second-order Bessel function, and $\lambda = 2\pi F b / B^2$.
We distinguish the two amplitudes $v_r$ and $a_r$ when necessary.

In the case of a purely two-dimensional (2D) electronic structure, the FS would be a straight cylinder showing a single dHvA frequency $F$, and $F\cos\theta$ would be constant irrespective of the field direction $\theta$.
With a quadratic in-plane dispersion, $m^*\cos\theta$ would also be constant; i.e., $m^*\cos\theta=m_0^*$, $m_0^*$ being the effective mass for $\theta=0$.
Given the crystal structure of KFe$_2$As$_2$, each FS cylinder would contribute 
\begin{equation}
\gamma = 1.452 \mu^*_0 \qquad \mathrm{(mJ mol}^{-1}\mathrm{K}^{-2}\mathrm{)}
\end{equation}
to the Sommerfeld coefficient $\gamma$ of the specific heat, where $\mu^*_0=m_0^*/m_e$.
Note that the contribution does not depends on the size of the FS cylinder.
\footnote{For a 2D system, the number of states is given by $N=(V/4\pi^3)A_0t$, where $A_0$ is the FS cross section normal to the $c$ axis and $t$ is the thickness of the Brillouin zone along the $c$ axis.  Since $m^*_0=(\hbar^2/2\pi)\partial A_0/\partial E$, the density of states $D=\partial N/\partial E$ is proportional to $m_0^*$ but does not depend on the FS size.}

In reality, there is a $c$-axis dispersion of the electronic band energy, which leads to corrugation of the cylindrical FS and produces at least two dHvA frequencies corresponding to the maximal and minimal FS cross sections.
Yamaji considered the simplest case where the $c$-axis dispersion takes a form of $\cos I_c k_z$, where $I_c$ is the interlayer distance and equals $c/2$ in KFe$_2$As$_2$, and derived the angular dependence of the two frequencies:\cite{Yamaji89JPSJ}
\begin{eqnarray}
F_{\pm}\cos\theta = F_0 \pm \frac{\Delta F_0}{2} J_0(I_ck_F \tan \theta),
\end{eqnarray}
where $F_0$ and $\Delta F_0$ are the average of and the difference between the two frequencies at $\theta$ = 0, respectively, $J_0$ is the zeroth-order Bessel function, and $k_F$ is the in-plane Fermi wave number.
At magic angles where $I_ck_F\tan\theta = \pi (n-\psi)$ ($n$: integer), the two frequencies coincide, resulting in single-frequency dHvA oscillations with an enhanced amplitude.
It is also known that the interlayer magnetoresistance at a constant field shows maxima at these field angles as a function of the field direction when the field is tilted from the $c$ axis.
This phenomenon is called angle dependent magnetoresistance oscillation (AMRO) and is used to determine the FS in quasi-2D metals.
The phase $\psi$ is 1/4 for the above simplest dispersion but actually varies depending on details of the $c$-axis dispersion.

\section{Results}

\subsection{Upper critical field $B_{c2}$}

\begin{figure}
\includegraphics[width=8cm]{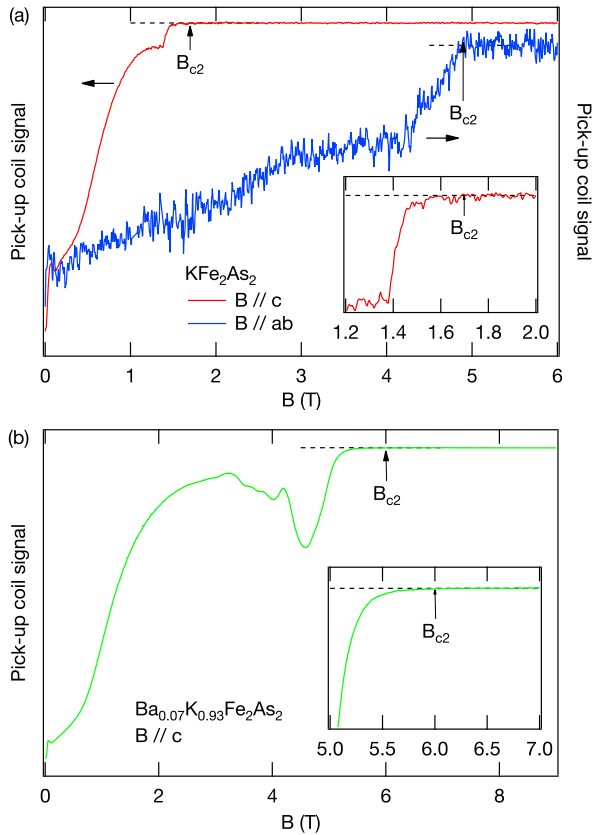}
\caption{\label{Xac}(color online)
Pick-up coil signals detected at the modulation frequency as a function of magnetic field for KFe$_2$As$_2$ (a) and Ba$_{0.07}$K$_{0.93}$Fe$_2$As$_2$ (b).  Upper critical fields $B_{c2}$ are indicated.  These field sweeps were made on cooling the dilution refrigerator.  The temperatures when $B = B_{c2}$ were 0.17 and 0.18 K for $B \parallel c$ and $B \parallel ab$ in (a) and 0.22 K in (b).}
\end{figure}

Figure~\ref{Xac}(a) shows pick-up coil signals detected at the modulation frequency $f$ as a function of magnetic field for KFe$_2$As$_2$.\footnote{Pick-up coil signals detected at the modulation frequency are proportional to ac magnetic susceptibilities for $B \parallel c$.  In the case of $B \parallel ab$, the coil axis is perpendicular to $B$, since the coil and sample are rotated together with the $c$ axis kept parallel to the coil axis.  If the alignment of the sample and coil axis is perfect, no signal will be detected.  However, there is small misalignment, which produces small signal proportional to the component of ac magnetization along the coil axis for $B \parallel ab$.} 
From these data, we estimate that the upper critical field $B_{c2}$ in this sample is 1.7 T at $T$ = 0.17 K for $B \parallel c$ and 4.9 T at $T$ = 0.18 K for $B \parallel ab$.
These values are slightly larger than those previously determined for a lower-quality sample.\cite{Terashima09JPSJKFA}
We note that the $ab$-plane critical field, which is consistent with Ref. \onlinecite{Burger13condmat}, is much lower than the orbital critical field ($\sim$7 T) estimated from the initial slope of $B_{c2}$ at $T=T_c$ in samples of similar quality to the present one,\cite{Kim_C11PRB, Abdel-Hafiez12PRB} confirming the existence of strong spin paramagnetic effects.\cite{Terashima09JPSJKFA}

Figure~\ref{Xac}(b) shows pick-up coil signal for $B \parallel c$ in Ba$_{0.07}$K$_{0.93}$Fe$_2$As$_2$ as a function of field, from which $B_{c2}$ in this sample is estimated to be 6.0 T at $T$ = 0.22 K for $B \parallel c$.
A more detailed study on $B_{c2}$ in Ba$_{0.07}$K$_{0.93}$Fe$_2$As$_2$ is given elsewhere.\cite{Terashima13PRB}

\subsection{dHvA oscillations}

\begin{figure}
\includegraphics[width=8cm]{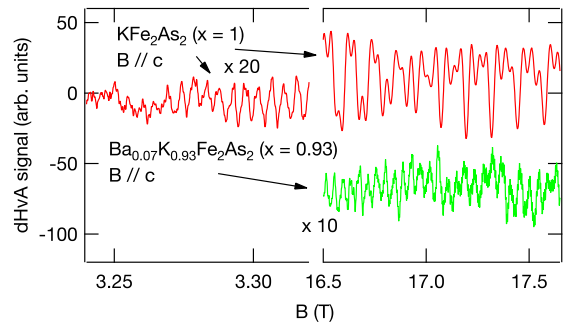}
\caption{\label{Signals}(color online)
Examples of dHvA oscillations for KFe$_2$As$_2$ (upper) and Ba$_{0.07}$K$_{0.93}$Fe$_2$As$_2$ (lower).  $T < 0.1$ K.}
\end{figure}

Figure~\ref{Signals} shows examples of dHvA oscillations.
For KFe$_2$As$_2$, the frequency of prominent oscillations for $B \parallel c$ is about 2.4 kT both at low fields and at high fields, though higher frequencies are also noticeable at high fields.
The oscillations continue down to below 3 T.\cite{Terashima10PRL_comment}
For Ba$_{0.07}$K$_{0.93}$Fe$_2$As$_2$, the frequency of prominent oscillations is about 6.7 kT.

\begin{figure}
\includegraphics[width=8cm]{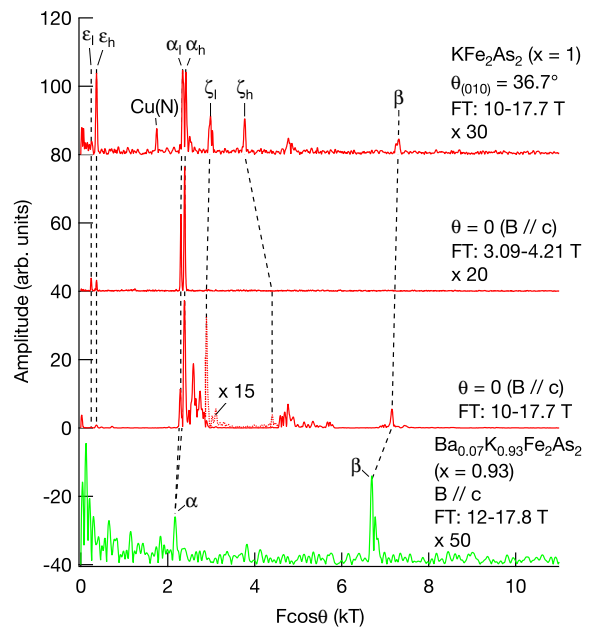}
\caption{\label{ExampleFTs}(color online)
Examples of Fourier transforms in $1/B$ of dHvA oscillations for KFe$_2$As$_2$ (upper three spectra) and Ba$_{0.07}$K$_{0.93}$Fe$_2$As$_2$ (lowest).  Note that the horizontal axis is $F\cos\theta$.  Fundamental frequencies are labeled with Greek letters.  The frequency marked Cu(N) is assigned to the copper neck oscillation from copper wire of the pick-up coil.}
\end{figure}

\begin{figure*}
\includegraphics[width=13.5cm]{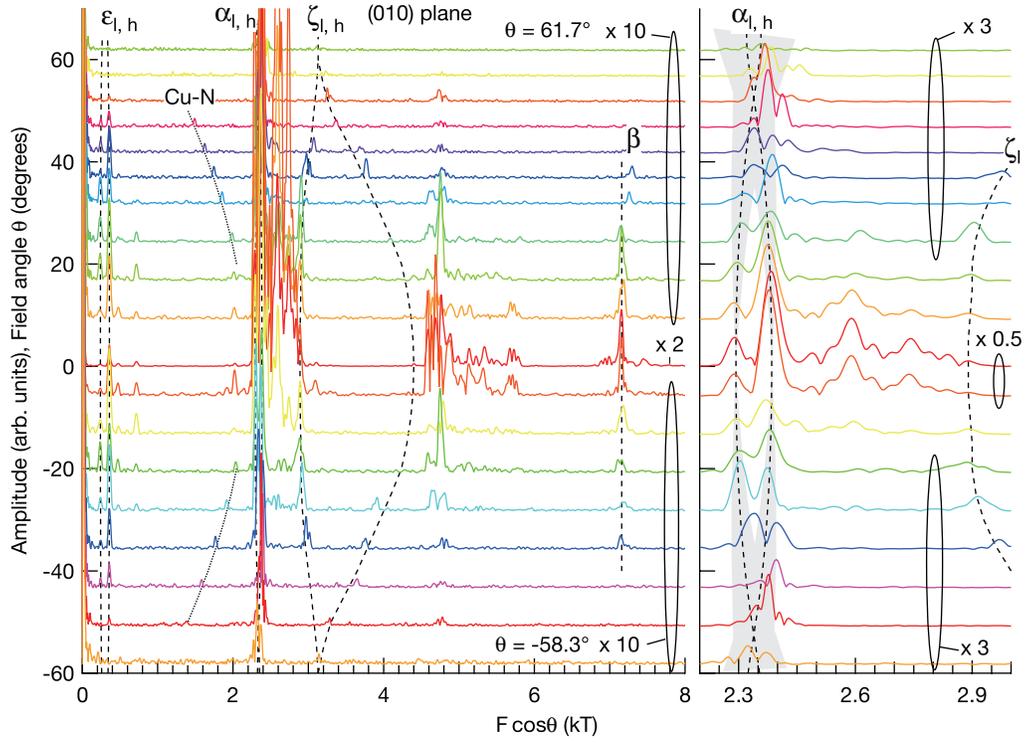}
\caption{\label{FTs010}(color online)
Angular variation of Fourier transforms of dHvA oscillations in KFe$_2$As$_2$ for the (010) plane.  The used field window is between 10 and 17.65 T.  The horizontal axis is $F\cos\theta$.  The spectra are shifted vertically so that the baseline of a spectrum measured at $\theta$ degrees is set at $\theta$.  The dashed lines labelled $\epsilon_{l, h}$ and $\alpha_{l, h}$ are based on the Yamaji model (see text), while those labelled $\zeta_{l, h}$ and $\beta$ are guides to the eye.  The right panel shows a frequency region near the $\alpha_{l, h}$ frequencies in an expanded scale.  The shading indicates a frequency spread of $\alpha_{l, h}$ calculated for a $\pm1^{\circ}$ error or distribution of the $c$ axis.}
\end{figure*}

\begin{figure*}
\includegraphics[width=13.5cm]{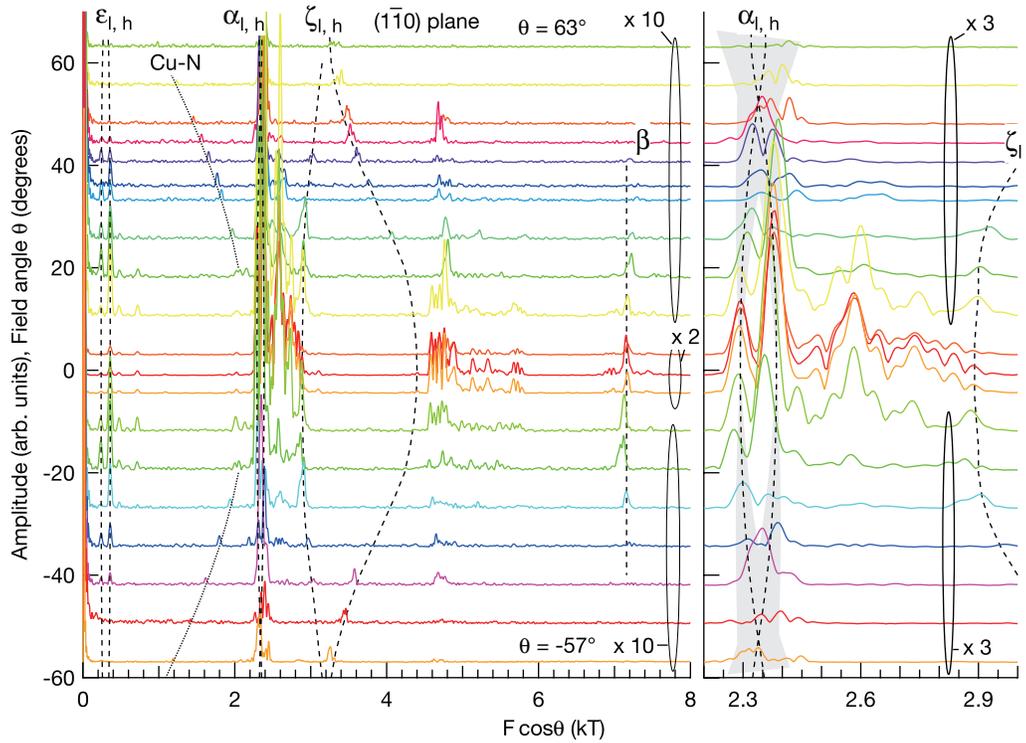}
\caption{\label{FTs110}(color online)
Same as Fig.~\ref{FTs010}, but for the (1$\bar{1}$0) plane.}
\end{figure*}

\begin{figure*}
\includegraphics[width=17.2cm]{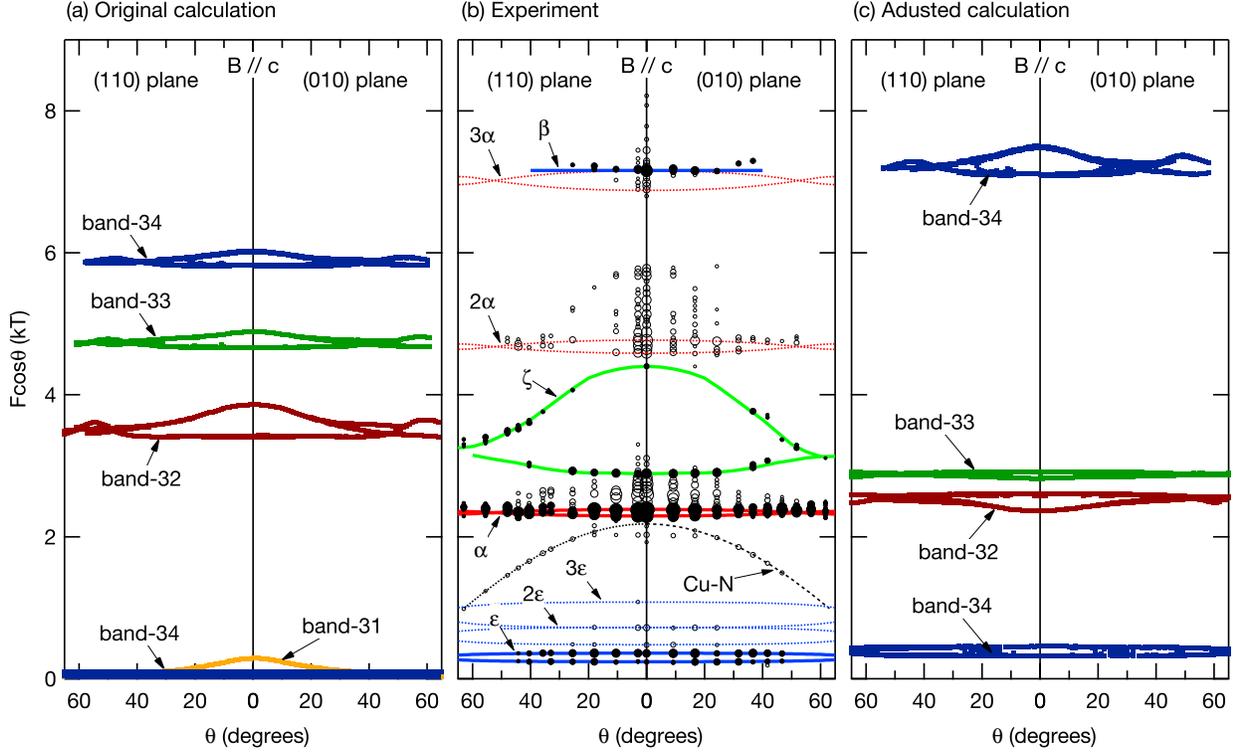}
\caption{\label{FvsAng}(color online)
Angular variation of calculated [(a) and (c)] and experimental dHvA frequencies (b).  Note that the vertical axis is $F\cos\theta$.  For the calculated frequencies, both original (a) and adjusted ones (c) are shown (see text).  For the experimental frequencies (b), the mark sizes are based on the oscillation amplitudes logarithmically.  Frequencies assigned to fundamentals are shown by filled marks.  The solid curves indicating fundamental frequencies in (b) are the same as the dashed lines in Figs.~\ref{FTs010} and \ref{FTs110}.  The dotted curves indicate positions of harmonic frequencies calculated from the corresponding solid curves.  The dashed line labelled Cu-N indicates the copper neck frequency arising from copper wire of the pick-up coil.}
\end{figure*}

\begin{figure*}
\includegraphics[width=15cm]{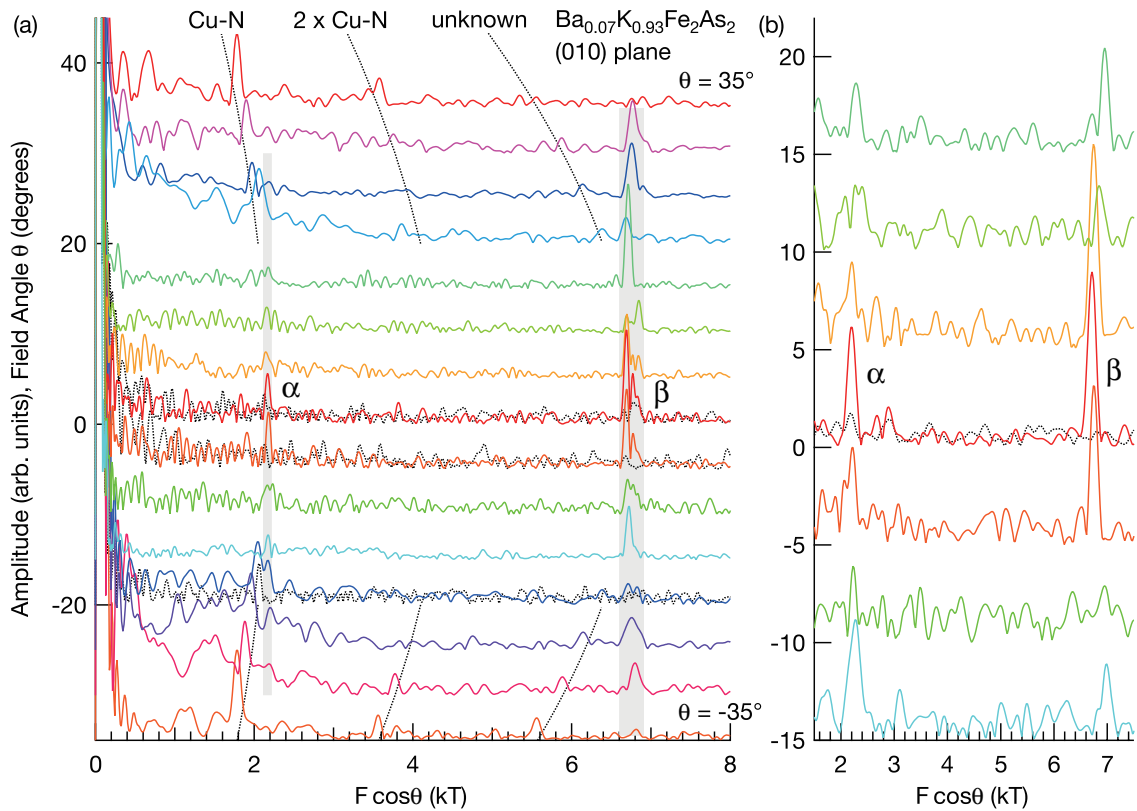}
\caption{\label{FTs093}(color online)
(a) Angular variation of Fourier transforms (solid lines) of dHvA oscillations in Ba$_{0.07}$K$_{0.93}$Fe$_2$As$_2$ for the (010) plane.  The field window is between 12 and 17.75 T for $|\theta| \leq 15^{\circ}$ and between 14 and 17.75 T for $|\theta| \geq 20^{\circ}$.  The horizontal axis is $F\cos\theta$.  The spectra are shifted vertically so that the baseline of a spectrum measured at $\theta$ degrees is set at $\theta$.  Three frequencies due to the pick-up coil (not the sample) are also observed as indicated: copper neck frequency (Cu-N), its second harmonic (2 $\times$ Cu-N), and an unknown frequency, which might be due to tin from solder used for wiring. These frequencies have detectable amplitudes only for $|\theta| \gtrsim 20^{\circ}$.  The spectra shown by dotted lines for some angles are Fourier transforms of pick-up coil signals \textit{without a sample} at the same angles.  The comparison indicates that the $\alpha$ and $\beta$ frequencies are not from the pick-up coil but from the sample.  (b) Fourier transforms for the field window between 12 and 14 T.  Although the frequency resolution deteriorates because of the narrower field window, the $\alpha$ frequency shows up slightly clearer than in (a).}
\end{figure*}

Figure~\ref{ExampleFTs} shows representative Fourier transforms of dHvA oscillations for the two compounds.
Fundamental frequencies, $\epsilon_{(l, h)}$, $\alpha_{(l, h)}$, $\zeta_{(l, h)}$, and $\beta$, are indicated.
The identification of these fundamentals is explained below.
We note that the $\alpha$ and $\beta$ frequencies in Ba$_{0.07}$K$_{0.93}$Fe$_2$As$_2$ are about 93\% of those in KFe$_2$As$_2$, which is consistent with the carrier number expected from the composition $x$ = 0.93.
Figures~\ref{FTs010} and \ref{FTs110} show details of angular variation of dHvA oscillations in KFe$_2$As$_2$ for fields in the (010) and the (1$\bar{1}$0) plane, respectively.
The angle dependences of the observed frequencies are shown in Fig.~\ref{FvsAng}(b).
Figure~\ref{FTs093} shows angular variation of dHvA oscillations in Ba$_{0.07}$K$_{0.93}$Fe$_2$As$_2$ for fields in the (010).
In the case of Ba$_{0.07}$K$_{0.93}$Fe$_2$As$_2$, three frequencies arising from materials of the pick-up coil assembly have comparable amplitudes to the frequencies from the sample: i.e., copper neck frequency, its second harmonic, and an unknown frequency, which might be due to tin from solder used for wiring.
However, they have detectable amplitudes only for $|\theta| \gtrsim 20^{\circ}$.\footnote{This is related to the imbalance of the pick-up coil.  The pick-up coil is composed of coaxially wound inner and outer coils.  The two coils are balanced without a sample at $\theta$ = 0 so that the emf's induced in the two coils cancel each other out.  The balance however degrades as $\theta$ increases.}
We show Fourier transforms of pick-up coil signals \textit{without a sample} for some field directions for comparison (dotted spectra), which clearly indicate that the $\alpha$ and $\beta$ frequencies arise from the sample, not from the coil assembly.

\begin{figure}
\includegraphics[width=8cm]{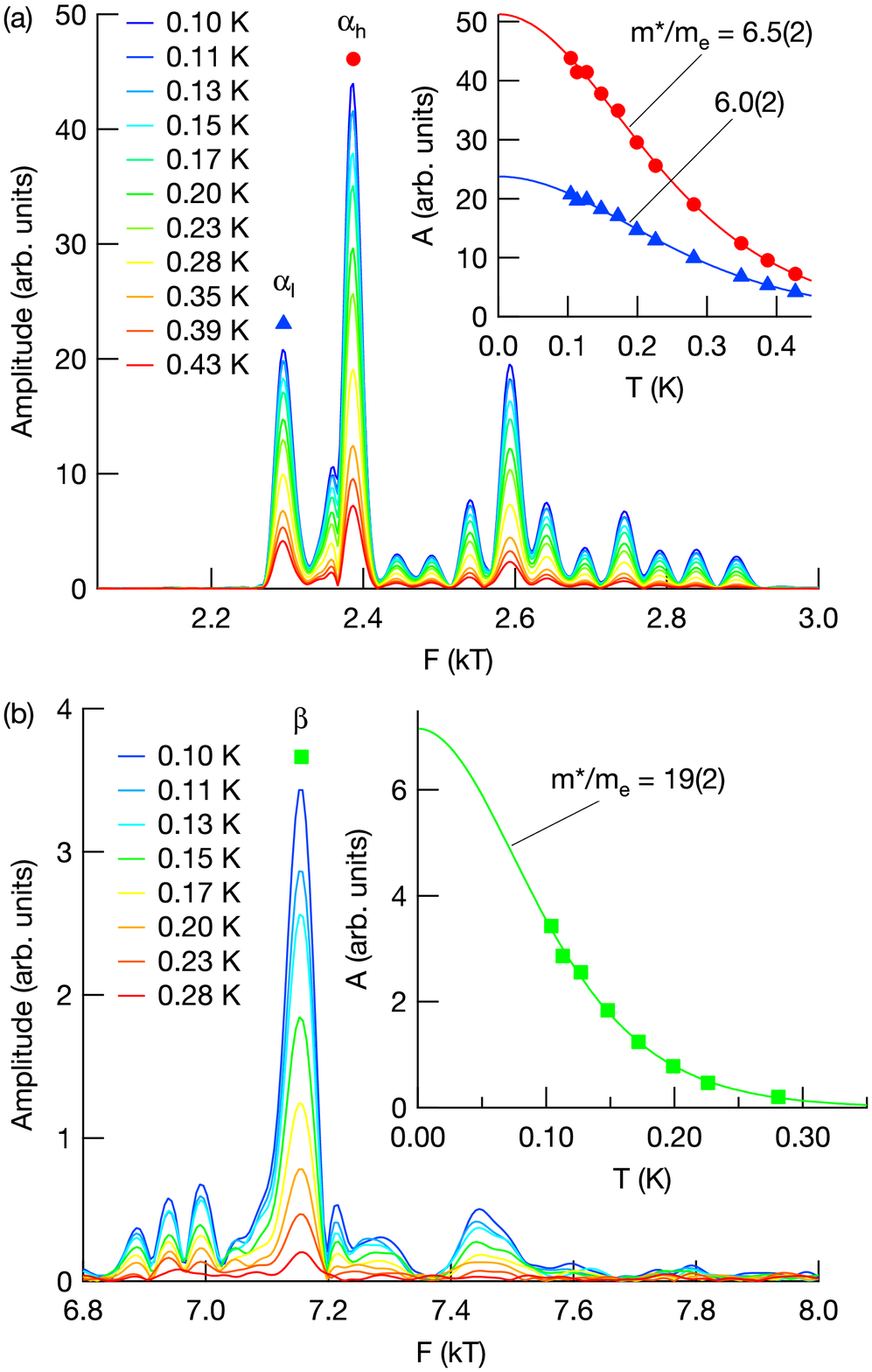}
\caption{\label{mass}(color online)
Temperature variation of (a) $\alpha_l$, $\alpha_h$, and (b) $\beta$ Fourier peaks for $B \parallel c$ in KFe$_2$As$_2$.  The insets show their amplitudes as functions of temperature.  The solid curves are fits to the Lifshitz-Kosevich formula, from which effective masses are estimated.}
\end{figure}

\begin{figure}
\includegraphics[width=8cm]{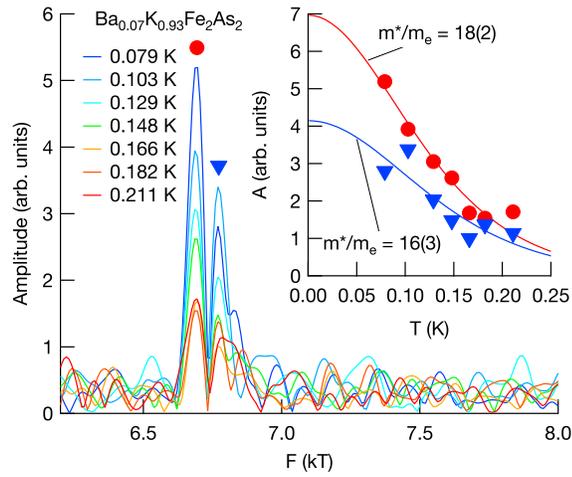}
\caption{\label{mass093}(color online)
Same as Fig.~\ref{mass}, but for $\beta$ peaks in Ba$_{0.07}$K$_{0.93}$Fe$_2$As$_2$.}
\end{figure}

Figures~\ref{mass} and \ref{mass093} exemplify determination of effective masses for KFe$_2$As$_2$ and Ba$_{0.07}$K$_{0.93}$Fe$_2$As$_2$, respectively.
The determined masses are tabulated in Tables I and II.
Heavy masses up to 19 $m_e$ for $B \parallel c$ are observed for KFe$_2$As$_2$.
The masses in Ba$_{0.07}$K$_{0.93}$Fe$_2$As$_2$ are nearly the same as the corresponding masses in KFe$_2$As$_2$.

\begin{table}%[H] add [H] placement to break table across pages
\caption{\label{Tab1}Experimental dHvA frequencies, effective masses, and Dingle temperatures in KFe$_2$As$_2$.}
\begin{ruledtabular}
\begin{tabular}{ccccc}
Field direction & Branch & $F$ (kT) & $m^*/m_e$ & $x_D^*$ (K)\\
\hline
$B \parallel c$ & $\epsilon_l$ & 0.24 & 6.0(4) & 0.14(2)\\
 & $\epsilon_h$ & 0.36 & 7.2(2) & 0.18(1)\\
 & $\alpha_l$ & 2.30 & 6.0(2) & 0.19(2)\\
 & $\alpha_h$ & 2.39 & 6.5(2) & 0.18(2)\\
 & $\zeta_l$ & 2.89 & 8.5(2) & $\sim$0.1\\
 & $\zeta_h$ & 4.40 & 18(2)&\\
 & $\beta$ & 7.16 & 19(2)&\\
 \\
$\theta_{(010)} = 36.7^\circ$ & $\epsilon_l$ & 0.30 & 7.4(7) &\\
 & $\epsilon_h$ & 0.45 & 8.4(2)&\\
 & $\alpha_l$ & 2.91 & 7.4(2) &\\
 & $\alpha_h$\footnotemark[1] & 2.98 & 7.8(2) &\\
 & $\alpha_h$\footnotemark[1] & 3.02 & 7.4(2) &\\
 & $\zeta_l$ & 3.72 & 11.1(4) &\\
 & $\zeta_h$ & 4.70 & 12.5(7) &\\
 & $\beta$ & 9.12 & 20(3) &\\
 \\
$\theta_{(1\bar{1}0)} = 40.5^\circ$ & $\epsilon_l$ & 0.32 & 7.7(7) &\\
 & $\epsilon_h$ & 0.47 & 9(2) &\\
 & $\alpha_l$ & 3.05 & 6.9(3) &\\
 & $\alpha_h$ & 3.13 & 7.7(2) &\\
 & $\zeta_l$\footnotemark[1] & 3.95 & 11(1) &\\
 & $\zeta_l$\footnotemark[1] & 3.99 & 13(1) &\\
 & $\zeta_h$ & 4.75 & 11(1) &\\
 & $\beta$ & unobserved & &\\
% Lines of table here ending with \\
\end{tabular}
\end{ruledtabular}
\footnotetext[1]{Extra splitting observed.}
\end{table}

\begin{table}%[H] add [H] placement to break table across pages
\caption{\label{Tab2}Experimental dHvA frequencies and effective masses in Ba$_{0.07}$K$_{0.93}$Fe$_2$As$_2$.}
\begin{ruledtabular}
\begin{tabular}{ccccc}
Field direction & Branch & $F$ (kT) & $m^*/m_e$\\
\hline
$B \parallel c$ & $\alpha$ & 2.17 & 4(1) \\
 & $\beta_l$ & 6.69 & 18(2) \\
 & $\beta_h$ & 6.77 & 16(3) \\
\end{tabular}
\end{ruledtabular}
\end{table}

\subsection{Fundamental frequencies}

\begin{figure}
\includegraphics[width=8.6cm]{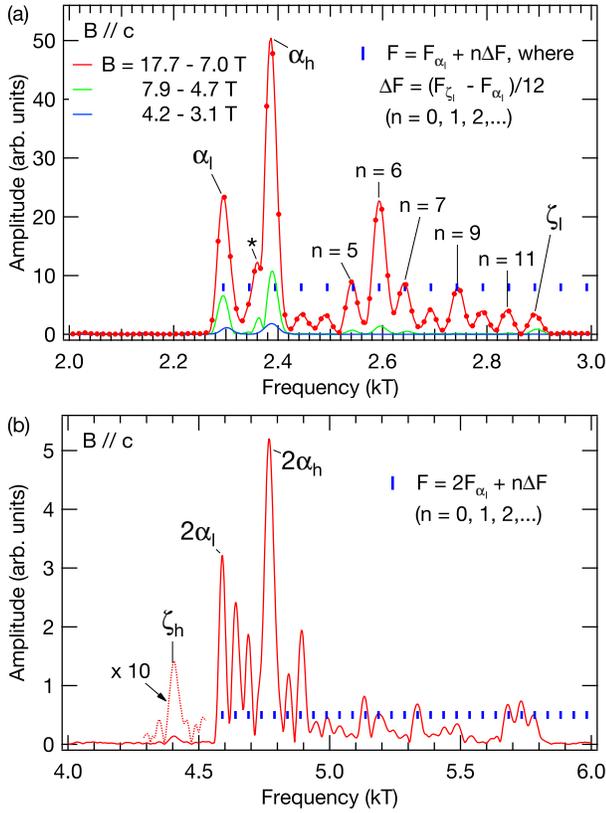}
\caption{\label{MB}(color online)
(a) Expanded view of Fourier transforms for $B \parallel c$ in KFe$_2$As$_2$ in a frequency range between 2 and 3 kT.   All the frequencies other than $\alpha_l$, $\alpha_h$, and $\zeta_l$ can be indexed with a formula $F = F_{\alpha_l}+n\Delta F$ ($n$: integer), where $\Delta F = (F_{\zeta_l}-F_{\alpha_l})/12$, as indicated by short vertical bars.  A small peak marked by an asterisk dose not accord with the formula.  It is however an artifact due to spectral interpolation by zero padding (the spectra shown by the solid curves are those interpolated by zero padding in the 1/$B$ domain).  The non-padded spectrum shown by dots can be consistent with a small $n$ = 1 peak buried under the low-frequency tail of the $\alpha_h$ peak.  Compare the three spectra for three different field windows: as the field is decreased, the peaks other than $\alpha_l$, $\alpha_h$, and $\zeta_l$ are quickly suppressed.  (b) Expanded view for a frequency range between 4 and 6 kT.  The frequencies higher than the second harmonic of $\alpha_l$, except for the second harmonic of $\alpha_h$, can be explained by $F = 2F_{\alpha_l}+n\Delta F$ ($n$: integer) as indicated by small vertical bars.  A small peak $\zeta_h$ at $F$ = 4.40 kT can not be explained and is a fundamental frequency.}
\end{figure}

\begin{figure}
\includegraphics[width=8cm]{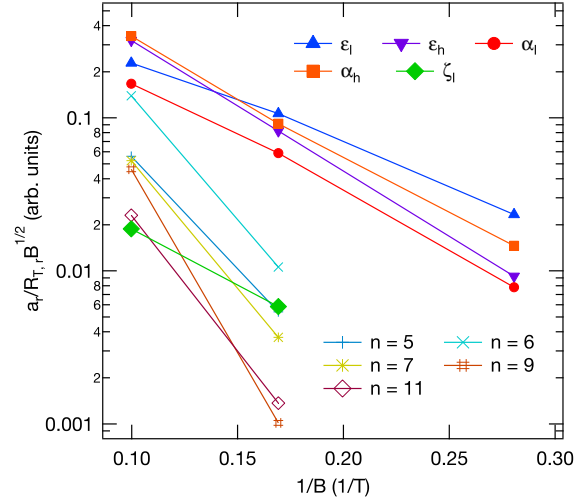}
\caption{\label{Ding}(color online)
Dingle plots for various frequencies in KFe$_2$As$_2$ for $B \parallel c$.  The frequencies indexed by $n$ are explained in Fig.~\ref{MB}.  According to the Lifshitz-Kosevich formula [see Eqs. (2)-(4)], this type of plot gives a straight line, the slope of which gives the Dingle temperature (Table I).  A very wide field range between 3.09 and 17.65 T was used to make these plots.  However,  in order to resolve finely spaced frequencies, wide field windows are necessary for Fourier transformation, and hence a very limited number of data points is obtained.  Note that frequencies other than $\alpha_l$, $\alpha_h$, and $\zeta_l$ are quickly suppressed as the field is decreased.}
\end{figure}

\begin{figure}
\includegraphics[width=6cm]{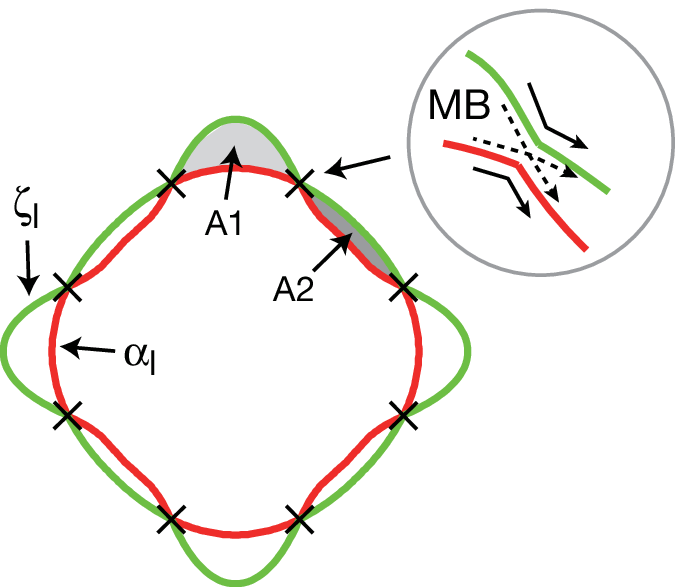}
\caption{\label{MBjunctions}(color online)
Schematic diagram of the $\alpha_l$ and $\zeta_l$ orbits deduced from recent ARPES data.\cite{Yoshida11JPCS, Wray12PRB, Yoshida12condmat}  Eight possible magnetic breakdown junctions, where holes can tunnel from the $\alpha_l$ orbit to $\zeta_l$ (or vice versa) through a small energy gap, and areas A1 and A2 are indicated.}
\end{figure}

A pair of frequencies $\epsilon_l$ and $\epsilon_h$ near $F\cos\theta$ = 0.3 kT in KFe$_2$As$_2$ [Figs.~\ref{FTs010}, \ref{FTs110}, and \ref{FvsAng}(b)] can be assigned to the minimum and maximum orbits on a corrugated FS cylinder.
Their angular variation is consistent with the Yamaji model as indicated in the figures.

A complex region between $F$ = 2 and 3 kT for $B \parallel c$ in KFe$_2$As$_2$ is shown in Fig.~\ref{MB}(a) in an expanded scale.
We identify three frequencies $\alpha_l$, $\alpha_h$, and $\zeta_l$ as fundamentals.
All the other frequencies can be indexed with a formula $F = F_{\alpha_l}+n\Delta F$ ($n$: integer), where $\Delta F = (F_{\zeta_l}-F_{\alpha_l})/12$.
Their amplitudes are quickly suppressed as the field is decreased [compare the three spectra for three different field windows in Fig.~\ref{MB}(a)].
To be more quantitative, Fig.~\ref{Ding} shows the field dependence of amplitudes of various frequencies for $B \parallel c$ in KFe$_2$As$_2$ in the form of Dingle plot.
The Dingle temperatures for fundamental frequencies estimated from the plots are shown in Table I.
(We can also estimate mean free paths of about 100 and 300 nm for $\epsilon$ and $\alpha$, respectively.) 
It is clear that the non-fundamental frequencies are suppressed with decreasing field more quickly than the fundamental frequencies.
These observations strongly suggest that the frequencies other than $\alpha_l$, $\alpha_h$, and $\zeta_l$ are due to magnetic breakdown orbits between $\alpha_l$ and $\zeta_l$.
We observed essentially the same spectra and field dependence in other samples.

%If the $\alpha_l$ and $\zeta_l$ orbits are at some point close enough for magnetic breakdown to occur, there will be four such points in the BZ in the simplest case due to the tetragonal symmetry.
%This can explain the $n$ = 3, 6, and 9 frequencies, and it is interesting to note that these frequencies are stronger than their neighboring ones [Fig.~\ref{MB}(a)].
%It is however unclear at present how the observed three-times finer splitting occurs.

Figure~\ref{MBjunctions} shows schematically the $\alpha_l$ and $\zeta_l$ orbits deduced from recent ARPES measurements.\cite{Yoshida11JPCS, Wray12PRB, Yoshida12condmat}
There are eight possible magnetic breakdown junctions as indicated.
If area A1 is approximately twice area A2 and hence $F_{A1} \approx 2\Delta F$ and $F_{A2} \approx \Delta F$, the observed magnetic breakdown frequencies can be explained:
frequency $n$ = 1 corresponds to $F_{\alpha_l}+F_{A2}$, $n$ = 2 to $F_{\alpha_l}+F_{A1}$ and $F_{\alpha_l}+2F_{A2}$, $n$ = 3 to $F_{\alpha_l}+F_{A1}+F_{A2}$ and $F_{\alpha_l}+3F_{A2}$, and so on.
It is interesting that the positions of the inferred magnetic breakdown junctions remind us of those of the octet-line nodes in the superconducting gap observed in a laser-ARPES study.\cite{Okazaki12Science}

Figure~\ref{MB}(b) shows a frequency region near the second harmonic of $\alpha_{l, h}$ in an expanded scale.
Many frequencies are observed above 2$\alpha_l$, and they (except 2$\alpha_h$) can be indexed with $F = 2F_{\alpha_l}+n\Delta F$ ($n$: integer).
However, a small peak at $F$ = 4.40 kT can not be indexed and we identify it as the counterpart of $\zeta_l$, namely $\zeta_h$.

Detailed angular dependence of $\alpha_{l, h}$ in KFe$_2$As$_2$ is shown in the right panels of Figs.~\ref{FTs010} and \ref{FTs110}. 
As the field is tilted from the $c$ axis, the two peaks gradually approach and roughly merge when $|\theta_{(010)}|$ is $\sim50^{\circ}$ or when $|\theta_{(1\bar{1}0)}|$ is slightly larger than 40$^{\circ}$.
The amplitudes at these magic angles are enhanced when compared to those at neighboring angles.
These observations are basically consistent with the Yamaji model (dashed lines).
Slight deviations of the peak positions from the model and extra splitting of the frequencies, which is especially noticeable at high angles $|\theta| \gtrsim 50^{\circ}$, can be explained by small error in $\theta$ and small distribution of the $c$-axis orientation in the sample.
The shading in the figures illustrates how the frequencies spread if there is a $\pm1^{\circ}$ error or distribution of the $c$-axis orientation. 
The magic angles also deviate from the Yamaji model and exhibit in-plane anisotropy; i.e, the magic angle in the (010) plane is slightly larger than that in the (1$\bar{1}$0) plane.
This indicates in-plane anisotropy of $k_F$ and/or more complex $c$-axis dispersion than the Yamaji model.
The observed magic angles and their anisotropy are consistent with those observed in AMRO measurements.\cite{Kimata10PRL}

The angular dependence of $\zeta_{l, h}$ [Figs.~\ref{FTs010}, \ref{FTs110}, and \ref{FvsAng}(b)] indicates that the $\zeta$ sheet is the most three-dimsnsional sheet of the Fermi surface.
The maximum frequency $\zeta_h$ exhibits much more variation in $F\cos\theta$ than the minimum frequency $\zeta_l$.
This indicates that, while the $\zeta$ cylinder is close to a straight one around the minimum cross section, it swells out locally around the maximum cross section.
For  $B \parallel c$, the effective mass of $\zeta_h$ is considerably larger than that of $\zeta_l$ (Table I).
As the field is tilted, the masses of the two frequencies become nearly the same (see the masses determined at $\theta_{(010)}$ = 36.7$^\circ$ and $\theta_{(1\bar{1}0)}$ = 40.5$^\circ$ in Table I).
This also indicates that the deformation of the $\zeta$ cylinder is local near the maximum cross section.

We now turn to the $\beta$ frequency in KFe$_2$As$_2$.
In the previous paper,\cite{Terashima10JPSJ} we identified this frequency with the third harmonic of $\alpha_h$ because it satisfies the relations $F_{\beta} = 3F_{\alpha_h}$ and $m^*_{\beta} = 3m^*_{\alpha_h}$ within experimental accuracy (Table I).
However, we have recently observed the $\beta$ frequency very clearly in Ba$_{0.07}$K$_{0.93}$Fe$_2$As$_2$ (Figs.~\ref{ExampleFTs} and \ref{FTs093}), where $F_{\beta}$ is unmistakably different from $3F_{\alpha_h}$.
The effective mass of $\beta$ in Ba$_{0.07}$K$_{0.93}$Fe$_2$As$_2$ is close to that in KFe$_2$As$_2$ (Tables I and II).
These findings have motivated us to reexamine the $\beta$ frequency in KFe$_2$As$_2$, and we have reached the conclusion that it is a fundamental frequency as explained below.

First of all, the amplitudes of the $\beta$ frequency appear too big for a third harmonic (Figs.~\ref{FTs010} and \ref{FTs110}).
According to the Lifshitz-Kosevich formula, amplitudes of harmonics basically decrease exponentially with the harmonic number $r$.
Because of the spin factor $R_{s, r}$, which oscillates with $r$, the third harmonic can accidentally be comparable to or even larger than the second harmonic for some field directions.
\footnote{Because $m^*$ varies as $\sim1/\cos\theta$, $S$ in Eq. (5) varies with $\theta$.} 
However, the $\beta$ frequency in KFe$_2$As$_2$ has a comparable amplitude to the second harmonic of $\alpha_h$ for a wide range of field directions, which seems difficult to explain if it is the third harmonic.

\begin{figure}
\includegraphics[width=8cm]{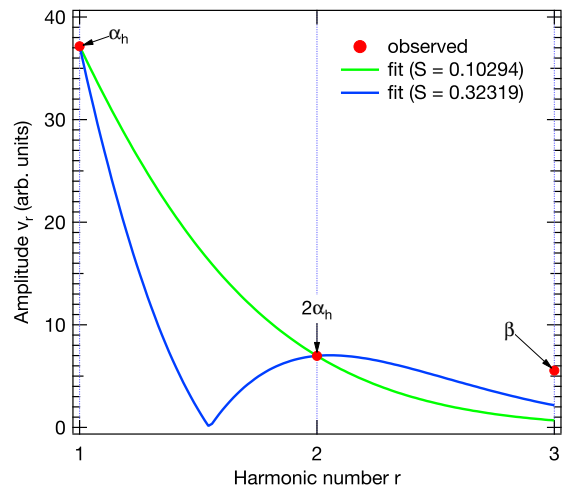}
\caption{\label{beta_amp}(color online)
Amplitudes of $\alpha_h$, its second harmonic, and $\beta$ for $B \parallel c$ in KFe$_2$As$_2$.  The solid curves are fits to the first two amplitudes based on the Lifshitz-Kosevich formula.  To illustrate the oscillatory nature of the spin factor $R_{s,r}$, those curves are shown as if the harmonic number $r$ is continuous.  The observed amplitude of $\beta$ is significantly larger than the amplitude of the third harmonic expected from either fit, indicating the existence of the $\beta$ fundamental frequency.}
\end{figure}

To be more quantitative, we analyze the amplitudes of $\alpha_h$, its second harmonic, and $\beta$ for $B \parallel c$, using Eqs. (1)-(6).
Since the effective mass $m^*$ and the Dingle temperature $x_D^*$ for the $\alpha_h$ frequency have been determined from the temperature and field dependences of the amplitude (Table I), we can determine the spin-splitting parameter $S$ from the amplitudes of the fundamental and second harmonic (Fig.~\ref{beta_amp}): $S=n\pm0.10294$ or $n\pm0.32319$ ($n$: integer).
Using these values of $S$, we can estimate the amplitude of the third harmonic.
In either case, the estimated amplitude is significantly smaller than the observed amplitude of the $\beta$ frequency.
It thus follows that, although there is some contribution from the third harmonic of $\alpha_h$, the observed amplitude of $\beta$ at $\theta$ = 0 is mostly due to the fundamental frequency $\beta$.

\begin{figure}
\includegraphics[width=8.6cm]{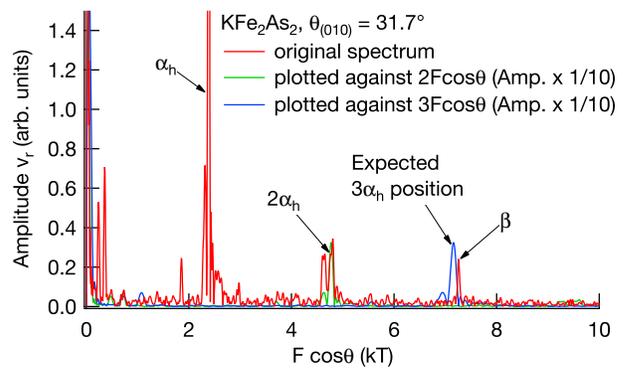}
\caption{\label{beta_3alpha_h}(color online)
Fourier transform spectrum at $\theta_{(010)} = 31.7^\circ$ for KFe$_2$As$_2$.  Two additional curves are shown: they are the same spectrum (with the amplitude reduced by 1/10) but plotted against 2Fcos$\theta$ and 3Fcos$\theta$, respectively, and hence indicate expected positions of second and third harmonics.  The $\beta$ peak clearly deviates from the expected position of the third harmonic of $\alpha_h$.}
\end{figure}

There is a further evidence for the fundamental $\beta$: the frequency of $\beta$ deviates from that of the third harmonic of $\alpha_h$ at high angles as shown in Fig.~\ref{beta_3alpha_h}.
We therefore conclude that the $\beta$ frequency is a fundamental frequency.

It is unclear at present whether the observed $\beta$ frequency is a maximum or a minimum frequency.
A small peak at $F$ = 7.44 kT at $\theta$ = 0 might perhaps be the counterpart (Figs.~\ref{FTs010} and \ref{FTs110}).

\begin{figure}
\includegraphics[width=8.6cm]{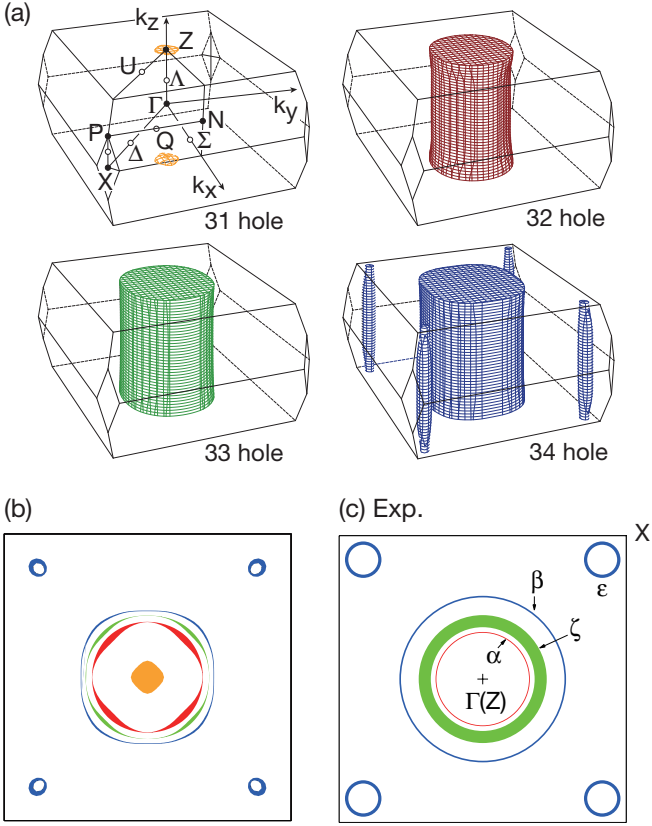}
\caption{\label{FScolored}(color online)
(a) Calculated Fermi surface and (b) its projection along the $c$ axis.  (No band-energy adjustments have been done.)  (c) Fermi surface cross-sections observed via the dHvA measurements.  The in-plane anisotropy is neglected.  The line thickness indicates the magnitude of the $c$-axis dispersion.}
\end{figure}

The experimentally determined Fermi surface in KFe$_2$As$_2$ is schematically shown in Fig.~\ref{FScolored}(c).

Lastly, we briefly discuss Ba$_{0.07}$K$_{0.93}$Fe$_2$As$_2$.
The minimum and maximum frequencies of $\alpha$ are not resolved in Fig.~\ref{FTs093}.
This is probably because the $\alpha$ oscillation is so weak that the quality of the signal is insufficient to resolve the two finely-spaced frequencies.
On the other hand, the $\beta$ frequency splits at $\theta$ = 0 (Fig.~\ref{FTs093} or Fig.~\ref{mass093}).
If the two frequencies are the minimum and maximum frequencies, it follows that the $\beta$ cylinder is strongly two dimensional ($\Delta F_0/F_0$ = 1.2\%).
The $\epsilon$ and $\zeta$ frequencies were not observed in Ba$_{0.07}$K$_{0.93}$Fe$_2$As$_2$.
This is probably because they are too weak.

The $\alpha$ and $\beta$ frequencies in Ba$_{0.07}$K$_{0.93}$Fe$_2$As$_2$ are both approximately 93\% of those in KFe$_2$As$_2$ and are consistent with the composition $x$ = 0.93 within experimental error.
This is compatible with a rigid band picture.
Within an effective mass approximation, $A=2\pi m_{band}E_F/\hbar^2$.
Hence, a dHvA frequency shift $\Delta F$ due to a shift in the Fermi energy $\Delta E_F$ satisfies a relation $\Delta F/F = \Delta E_F/E_F$.
Therefore, if two bands have the same $E_F$, $\Delta F/F$ for a energy shift $\Delta E_F$ is the same for both bands.
Assuming the same mass renormalization for both bands, $E_F \propto F/m_{band} \propto F/m^*$.
Since $F/(m^*/m_e)$ = 0.38 kT for both $\alpha$ and $\beta$ orbits in KFe$_2$As$_2$, $\Delta F/F$ should be the same as observed.

\section{Comparison to band structure calculations}

\begin{figure}
\includegraphics{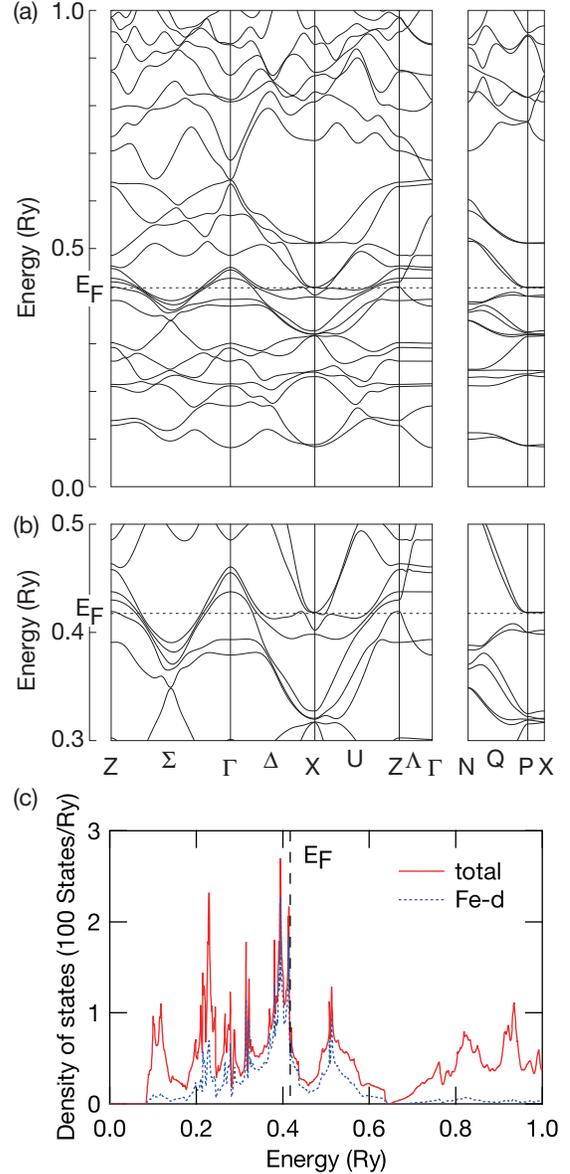}
\caption{\label{Band_and_DOS}(color online)
(a) Electronic band structure, (b) blowup of a region near $E_F$, and (c) density of states of KFe$_2$As$_2$ calculated with the spin-orbit interaction included.\cite{Terashima10JPSJ}  Points of symmetry (solid circles) and lines of symmetry (open circles) in the Brillouin zone are explained in the top left figure of Fig.~\ref{FScolored}(a).}
\end{figure}

%\begin{figure*}
%\includegraphics[width=17.2cm]{Band_calc}
%\caption{\label{band_calc}(color online)
%(a) Electronic band structure, (b) density of states, and (c) Fermi surface of KFe$_2$As$_2$ calculated with the spin-orbit interaction included.\cite{Terashima10JPSJ}  Points of symmetry (solid circles) and lines of symmetry (open circles) in the Brillouin zone are explained in the top left figure of (c).}
%\end{figure*}

The electronic band structure of KFe$_2$As$_2$ was calculated within the local density approximation (LDA) by using a full-potential linearized augmented plane-wave  (FLAPW) method.\cite{Terashima10JPSJ}
We used the program codes TSPACE\cite{Yanase1995} and KANSAI-06.
The experimental crystal structure \cite{Rozsa81ZNB} including the atomic position $z_{As}$ of As was used for the calculation.
The calculated Fermi surface is shown in Fig.~\ref{FScolored}, and the electronic band structure and density of states (DOS) are shown in Fig.~\ref{Band_and_DOS}.
The spin-orbit interaction has been included in these calculations.
The DOS at the Fermi level ($E_F$) is 58.4 states/(Ry f.u.), which corresponds to the Sommerfeld coefficient of $\gamma_{band}$ =  10.1 mJ/K$^2$mol.
Note that a direct specific-heat measurement gives $\gamma$ = 93  mJ/K$^2$mol,\cite{Fukazawa11JPSJ_SA} indicating a large mass enhancement of $\gamma/\gamma_{band}$ = 9 due to electronic correlations.
As can be seen from the DOS plot [Fig.~\ref{Band_and_DOS}(c)], the states near $E_F$ are mostly derived from the Fe 3$d$ orbitals.
Four bands 31--34 cross $E_F$ [Fig.~\ref{Band_and_DOS}(b)], and the FS consists of three concentric hole cylinders at the $\Gamma$ point of the BZ (bands-32, 33, and 34), small hole cylinders near the zone boundary (band-34), and a small hole pocket at Z (band-31) [Fig.~\ref{FScolored}(a)].
Note that the band-34 small hole cylinders are situated inside the Brillouin zone, not on the border, and hence there are four of them per Brillouin zone [Fig.~\ref{FScolored}(b)].
A previous calculation,\cite{Singh09PRB} where $z_{As}$ was relaxed by energy minimization, predicted an electron cylinder at X instead of the hole cylinders near X.
However, ARPES measurements so far have found no electron cylinder in KFe$_2$As$_2$ and are qualitatively consistent with the presently calculated FS.\cite{Sato09PRL, Yoshida11JPCS, Yoshida12condmat}

For the observed dHvA frequencies, it is clear that the $\alpha$, $\zeta$, and $\beta$ cylinders correspond to the three hole cylinders at $\Gamma$ and the $\epsilon$ cylinder to the small hole cylinder near X (Fig.~\ref{FScolored}).
However, the quantitative agreement is poor as can be seen in Fig.~\ref{FvsAng}, where the calculated and experimental frequencies are shown.
Calculated band masses ranging from 0.3 to 2.9 $m_e$ for $B \parallel c$ are much smaller than the observed effective masses (Table I), again suggesting the strong correlations. 

To see if the agreement can be improved, we have adjusted the band energies so that the smallest band-32 cylinder matches the $\alpha$ cylinder and that the largest band-34 cylinder matches the $\beta$ cylinder. 
The energy of band-33 was adjusted to keep the total carrier number.
The small band-31 pocket was neglected.
The resultant frequencies are shown in Fig.~\ref{FvsAng}(c).
The large undulation of the $\zeta$ cylinder can not be reproduced by the calculation.
\footnote{The adjusted calculation gives a Sommerfeld coefficient of 33.1 mJ/K$^2$mol.  However, the validity of this estimate is questioned because the adjustment can not reproduce the observed FS.} 

 %Dynamical mean field theory (DMFT) studies have shown that the crystal field splitting of the Fe $3d$ levels may be modified if electronic correlations are treated beyond the LDA.\cite{Haule08PRL, Aichhorn09PRB}
%The Fermi surface calculated in Ref.~\onlinecite{Yin11NatMat} using the DMFT has a large cylinder of $xy$ character and two similarly-sized smaller cylinders of $xz$/$yz$ character at the $\Gamma$ point.
%This is consistent with the above view.
%The failure of LDA calculations in KFe$_2$As$_2$ contrasts sharply with the case of BaFe$_2$As$_2$, where LDA calculations are basically correct.\cite{Terashima11PRL}

\section{Discussion}
The failure of LDA calculations in KFe$_2$As$_2$ contrasts sharply with the case of BaFe$_2$As$_2$, where LDA calculations give basically correct description of the Fermi surface (despite the overestimation of the antiferromagnetic moment).\cite{Terashima11PRL}
It seems that LDA band structure calculations fail to predict crystal field splitting of the Fe $3d$ levels in KFe$_2$As$_2$.
In the calculated band structure, one band is situated immediately above $E_F$ at the $\Gamma$ point and two nearly degenerate bands are situated above it [Fig.~\ref{Band_and_DOS}(b)].
The former band is of $xy$ character.
The latter two bands are of $xz$/$yz$ character and would be truly degenerate if spin-orbit coupling were absent.
However, experimental observation is that there are two similarly sized relatively small cylinders ($\alpha$ and $\zeta$)  and one relatively large cylinder ($\beta$) at the $\Gamma$ point.
This strongly suggests that the $xz$/$yz$ bands are actually lower than the $xy$ band and produce the $\alpha$ and $\zeta$ cylinders.
In fact, these orbital characters have been confirmed by recent ARPES measurements.\cite{Okazaki12Science, Yoshida12condmat}

Dynamical mean field theory (DMFT) studies have shown that the crystal field splitting of the Fe $3d$ levels may be modified if electronic correlations are treated beyond the LDA.\cite{Haule08PRL, Aichhorn09PRB}
A recent DMFT calculation\cite{Yin11NatMat} suggests that the order of the $xy$ and $xz$/$yz$ bands at the $\Gamma$ point in KFe$_2$As$_2$ are inverted by electronic correlations.
However, it should also be noted that the crystal field splitting is affected by hybridization between the Fe 3$d$ and As 4$p$ as well as Fe 4$p$ states.
Given the well known problem of band structure calculations in iron pnictides that structural optimization for paramagnetic states can not predict the correct position of As, this hybridization may not be calculated very accurately within the LDA, which would affect the crystal field splitting of the Fe 3$d$ levels.
This point deserves further studies.

\begin{table*}%[H] add [H] placement to break table across pages
\caption{\label{Tab3}Estimated FS volume $A$ and Sommerfeld coefficient $\gamma$.  The former is estimated from the average of a maximum and a minimum frequency.  The latter is estimated from the average of the effective masses for the maximum and minimum frequencies within the 2D approximation [i.e., Eq. (7) with $m^*\cos\theta=m^*_0$].  Two sets of data, one at $B \parallel c$ and the other at $\theta_{(010)} = 36.7^\circ$, are separately used for the estimation.  FS volumes determined by ARPES\cite{Sato09PRL, Yoshida11JPCS} and AMRO\cite{Kimata10PRL} measurements are also shown.}
\begin{ruledtabular}
\begin{tabular}{cccccccc}
& \multicolumn{4}{c}{dHvA (present work)} & \multicolumn{2}{c}{ARPES} & AMRO \\
\cline{2-5} \cline{6-7}
& \multicolumn{2}{c}{$B \parallel c$} & \multicolumn{2}{c}{$\theta_{(010)}=36.7^\circ$}  & Ref.~\onlinecite{Sato09PRL} & Ref.~\onlinecite{Yoshida11JPCS} & Ref.~\onlinecite{Kimata10PRL} \\
\cline{2-3} \cline{4-5}
FS & $A$ (\%BZ) & $\gamma$ (mJ mol$^{-1}$K$^{-2}$) & $A$ (\%BZ) & $\gamma$ (mJ mol$^{-1}$K$^{-2}$) & $A$ (\%BZ) & $A$ (\%BZ) & $A$ (\%BZ) \\
\hline
$\epsilon$\footnotemark[1] & 1.1 $\times$ 4 & 9.6 $\times$ 4 & 1.1 $\times$ 4 & 9.1 $\times$ 4 & 2.5 & 2.1 $\times$ 4 & \\
$\alpha$ & 8.4 & 9.1 & 8.5 & 8.7 & 7 & 10.1 & $\sim$12\\
$\zeta$ & 13.0 & 19.2 & 12.1 & 13.7 &   & 11.8 & $\sim$17\\
$\beta$ & 25.6 & 27.6 & 25.5 & 23.2 & 22 $\times$ 2 & 28.5 & \\
\hline
total & 51.4 & 94 & 50.5 & 82 & 53.5 & 58.8 & \\
% Lines of table here ending with \\
\end{tabular}
\end{ruledtabular}
\footnotetext[1]{There are four $\epsilon$ cylinders in the BZ.}
\end{table*}

We can estimate the Fermi surface volume and Sommerfeld coefficient of the specific heat from the determined FS within a 2D approximation (Table III).
We have used two data sets, one at $\theta$=0 and the other at $\theta$ = 36.7$^{\circ}$, separately for the estimation, and have obtained consistent results.

The estimated FS volume corresponds to the carrier number of  1.01 -- 1.03 holes/f.u., which is consistent with the stoichiometry of KFe$_2$As$_2$ within experimental error.
We also compare the sizes of the observed FS cylinders with those obtained by ARPES\cite{Sato09PRL, Yoshida11JPCS} and AMRO\cite{Kimata10PRL} measurements in Table III.
The more recent ARPES data (Ref.~\onlinecite{Yoshida11JPCS}) are in reasonable agreement with the present data.
Usually AMRO measurements give quantitative estimates of FS sizes, but in the present case the agreement with the present data is limited, which is ascribed to the fact that only a small number of broad AMRO peaks were observed.\cite{Kimata10PRL}

The Sommerfeld coefficient is estimated to be 82 -- 94 mJmol$^{-1}$K$^{-2}$, which is consistent with the direct measure of $\gamma$ = 93 mJmol$^{-1}$K$^{-2}$ (Ref. \onlinecite{Fukazawa11JPSJ_SA} and also Ref. \onlinecite{Hardy13condmat}) and also with an estimate of $\gamma \sim$ 84 mJmol$^{-1}$K$^{-2}$ in a recent ARPES study,\cite{Yoshida12condmat} where effective masses have been determined from band dispersions near the Fermi level. 
These confirm a mass enhancement of about 9 in KFe$_2$As$_2$.
Therefore, the electronic correlations are much stronger in KFe$_2$As$_2$ than in BaFe$_2$As$_2$, where mass enhancements are 2 -- 3.

It is interesting to recall here that overall band-width renormalization factors estimated in early ARPES studies do not vary significantly with the chemical composition in the Ba$_{1-x}$K$_x$Fe$_2$As$_2$ system:
1.5 in BaFe$_2$As$_2$,\cite{Yi09PRB} 2.7 in Ba$_{0.6}$K$_{0.4}$Fe$_2$As$_2$,\cite{Yi09PRB} and 2 in KFe$_2$As$_2$.\cite{Sato09PRL}
On the theoretical side, a study based on the fluctuation exchange (FLEX) approximation has predicted a significant growth in the mass enhancement from BaFe$_2$As$_2$ to KFe$_2$As$_2$,\cite{Ikeda10PRB} while the DMFT study of Ref.~\onlinecite{Yin11NatMat} suggests similar mass enhancements for both BaFe$_2$As$_2$ and KFe$_2$As$_2$.
These seemingly contradicting results can be reconciled if one notices the existence of two distinct mechanisms contributing to the mass enhancement as previously pointed out.\cite{Ortenzi09PRL}
One is narrowing of overall band widths, which can be measured as band-width renormalization factors in ARPES measurements and seems to be dealt with a DMFT fairly well.
The other is flattening of bands near the Fermi level, which arises from interaction with low-energy bosonic excitations.
The large discrepancy between the dHvA mass enhancement and the ARPES band-width renormalization factor in KFe$_2$As$_2$ indicates that this latter effect, which involves spin and orbital fluctuations, grows from BaFe$_2$As$_2$ to KFe$_2$As$_2$ as suggested by the FLEX study.\cite{Ikeda10PRB}
The existence of strong spin fluctuations in KFe$_2$As$_2$ has been evidenced by inelastic neutron scattering\cite{Lee11PRL} and NMR measurements.\cite{Zhang10PRB, Hirano12JPSJ}

\section{Summary}
We have performed dHvA measurements on KFe$_2$As$_2$ and Ba$_{0.07}$K$_{0.93}$Fe$_2$As$_2$.
For KFe$_2$As$_2$, we have identified frequencies $\epsilon$, $\alpha$, $\zeta$, and $\beta$ as fundamentals.
The rest of observed frequencies are attributed to harmonics and magnetic breakdown orbits between $\alpha_l$ and $\zeta_l$.
The positions of the inferred magnetic breakdown junctions between $\alpha_l$ and $\zeta_l$ resemble those of the octet-line nodes of the superconducting energy gap.\cite{Okazaki12Science} 
With the aid of LDA band structure calculations, we have assigned the $\epsilon$ frequency to a hole FS cylinder near the X point and the $\alpha$, $\zeta$, and $\beta$ frequencies to three concentric hole cylinders at the $\Gamma$ point.
On a quantitative level, however, the agreement between the observed and calculated frequencies is poor.
This can be attributed to the fact that the crystal field splitting of the Fe 3$d$ levels are not correctly calculated.
Effective masses are large, up to 19 $m_e$ for $B \parallel c$.
From the dHvA data, we estimate the carrier concentration and Sommerfeld coefficient to be 1.01 -- 1.03 holes/f.u. and 82 -- 94 mJmol$^{-1}$K$^{-2}$, respectively.
They are consistent with the chemical stoichiometry and a direct measure of 93 mJmol$^{-1}$K$^{-2}$,\cite{Fukazawa11JPSJ_SA} respectively, establishing that the determined FS is complete.
The large Sommerfeld coefficient, which is about 9 times larger than a band value, indicates that KFe$_2$As$_2$ is a strongly correlated metal.
The discrepancy between the mass enhancement of 9 and the ARPES overall band-width renormalization of 2 (Ref. \onlinecite{Sato09PRL}) suggests the importance of low-energy spin and/or orbital fluctuations.
For Ba$_{0.07}$K$_{0.93}$Fe$_2$As$_2$, we have observed the $\alpha$ and $\beta$ frequencies.
The frequency change between the two compounds can be understood within a rigid-band model.

\begin{acknowledgments}
TT thanks T. Yoshida for valuable discussions; especially, he pointed out possible relation between the magnetic breakdown junctions and octet-line nodes.
This work was supported by Grants-in-Aid for Scientific Research (Nos. 21540351, 22540380, and 22684016) from MEXT and JSPS, Japan, Grants-in-Aid for Scientific Research on Innovative Areas ''Heavy Electrons" (Nos. 20102005, 21102505, and 23102725) from MEXT, Japan, and Funding Program for World-Leading Innovative R\&D on Science and Technology (FIRST Program) from JSPS, Japan.
\end{acknowledgments}

%\bibliography{pnictides}

\begin{thebibliography}{66}%
\makeatletter
\providecommand \@ifxundefined [1]{%
 \@ifx{#1\undefined}
}%
\providecommand \@ifnum [1]{%
 \ifnum #1\expandafter \@firstoftwo
 \else \expandafter \@secondoftwo
 \fi
}%
\providecommand \@ifx [1]{%
 \ifx #1\expandafter \@firstoftwo
 \else \expandafter \@secondoftwo
 \fi
}%
\providecommand \natexlab [1]{#1}%
\providecommand \enquote  [1]{``#1''}%
\providecommand \bibnamefont  [1]{#1}%
\providecommand \bibfnamefont [1]{#1}%
\providecommand \citenamefont [1]{#1}%
\providecommand \href@noop [0]{\@secondoftwo}%
\providecommand \href [0]{\begingroup \@sanitize@url \@href}%
\providecommand \@href[1]{\@@startlink{#1}\@@href}%
\providecommand \@@href[1]{\endgroup#1\@@endlink}%
\providecommand \@sanitize@url [0]{\catcode `\\12\catcode `\$12\catcode
  `\&12\catcode `\#12\catcode `\^12\catcode `\_12\catcode `\%12\relax}%
\providecommand \@@startlink[1]{}%
\providecommand \@@endlink[0]{}%
\providecommand \url  [0]{\begingroup\@sanitize@url \@url }%
\providecommand \@url [1]{\endgroup\@href {#1}{\urlprefix }}%
\providecommand \urlprefix  [0]{URL }%
\providecommand \Eprint [0]{\href }%
\providecommand \doibase [0]{http://dx.doi.org/}%
\providecommand \selectlanguage [0]{\@gobble}%
\providecommand \bibinfo  [0]{\@secondoftwo}%
\providecommand \bibfield  [0]{\@secondoftwo}%
\providecommand \translation [1]{[#1]}%
\providecommand \BibitemOpen [0]{}%
\providecommand \bibitemStop [0]{}%
\providecommand \bibitemNoStop [0]{.\EOS\space}%
\providecommand \EOS [0]{\spacefactor3000\relax}%
\providecommand \BibitemShut  [1]{\csname bibitem#1\endcsname}%
\let\auto@bib@innerbib\@empty
%</preamble>
\bibitem [{\citenamefont {Kamihara}\ \emph {et~al.}(2008)\citenamefont
  {Kamihara}, \citenamefont {Watanabe}, \citenamefont {Hirano},\ and\
  \citenamefont {Hosono}}]{Kamihara08JACS}%
  \BibitemOpen
  \bibfield  {author} {\bibinfo {author} {\bibfnamefont {Y.}~\bibnamefont
  {Kamihara}}, \bibinfo {author} {\bibfnamefont {T.}~\bibnamefont {Watanabe}},
  \bibinfo {author} {\bibfnamefont {M.}~\bibnamefont {Hirano}}, \ and\ \bibinfo
  {author} {\bibfnamefont {H.}~\bibnamefont {Hosono}},\ }\href {\doibase
  10.1021/ja800073m} {\bibfield  {journal} {\bibinfo  {journal} {J. Am. Chem.
  Soc.}\ }\textbf {\bibinfo {volume} {130}},\ \bibinfo {pages} {3296} (\bibinfo
  {year} {2008})}\BibitemShut {NoStop}%
\bibitem [{\citenamefont {Ishida}\ \emph {et~al.}(2009)\citenamefont {Ishida},
  \citenamefont {Nakai},\ and\ \citenamefont {Hosono}}]{Ishida09JPSJ_review}%
  \BibitemOpen
  \bibfield  {author} {\bibinfo {author} {\bibfnamefont {K.}~\bibnamefont
  {Ishida}}, \bibinfo {author} {\bibfnamefont {Y.}~\bibnamefont {Nakai}}, \
  and\ \bibinfo {author} {\bibfnamefont {H.}~\bibnamefont {Hosono}},\ }\href
  {\doibase 10.1143/JPSJ.78.062001} {\bibfield  {journal} {\bibinfo  {journal}
  {J. Phys. Soc. Jpn.}\ }\textbf {\bibinfo {volume} {78}},\ \bibinfo {pages}
  {062001} (\bibinfo {year} {2009})}\BibitemShut {NoStop}%
\bibitem [{\citenamefont {Johnston}(2010)}]{Johnston10AdvPhys}%
  \BibitemOpen
  \bibfield  {author} {\bibinfo {author} {\bibfnamefont {D.~C.}\ \bibnamefont
  {Johnston}},\ }\href {\doibase 10.1080/00018732.2010.513480} {\bibfield
  {journal} {\bibinfo  {journal} {Adv. Phys.}\ }\textbf {\bibinfo {volume}
  {59}},\ \bibinfo {pages} {803} (\bibinfo {year} {2010})}\BibitemShut
  {NoStop}%
\bibitem [{\citenamefont {Kito}\ \emph {et~al.}(2008)\citenamefont {Kito},
  \citenamefont {Eisaki},\ and\ \citenamefont {Iyo}}]{Kito08JPSJ}%
  \BibitemOpen
  \bibfield  {author} {\bibinfo {author} {\bibfnamefont {H.}~\bibnamefont
  {Kito}}, \bibinfo {author} {\bibfnamefont {H.}~\bibnamefont {Eisaki}}, \ and\
  \bibinfo {author} {\bibfnamefont {A.}~\bibnamefont {Iyo}},\ }\href {\doibase
  10.1143/JPSJ.77.063707} {\bibfield  {journal} {\bibinfo  {journal} {J. Phys.
  Soc. Jpn.}\ }\textbf {\bibinfo {volume} {77}},\ \bibinfo {pages} {063707}
  (\bibinfo {year} {2008})}\BibitemShut {NoStop}%
\bibitem [{\citenamefont {Ren}\ \emph {et~al.}(2008)\citenamefont {Ren},
  \citenamefont {Lu}, \citenamefont {Yang}, \citenamefont {Yi}, \citenamefont
  {Shen}, \citenamefont {Li}, \citenamefont {Che}, \citenamefont {Dong},
  \citenamefont {Sun}, \citenamefont {Zhou},\ and\ \citenamefont
  {Zhao}}]{Ren08CPL}%
  \BibitemOpen
  \bibfield  {author} {\bibinfo {author} {\bibfnamefont {Z.-A.}\ \bibnamefont
  {Ren}}, \bibinfo {author} {\bibfnamefont {W.}~\bibnamefont {Lu}}, \bibinfo
  {author} {\bibfnamefont {J.}~\bibnamefont {Yang}}, \bibinfo {author}
  {\bibfnamefont {W.}~\bibnamefont {Yi}}, \bibinfo {author} {\bibfnamefont
  {X.-L.}\ \bibnamefont {Shen}}, \bibinfo {author} {\bibfnamefont {Z.-C.}\
  \bibnamefont {Li}}, \bibinfo {author} {\bibfnamefont {G.-C.}\ \bibnamefont
  {Che}}, \bibinfo {author} {\bibfnamefont {X.-L.}\ \bibnamefont {Dong}},
  \bibinfo {author} {\bibfnamefont {L.-L.}\ \bibnamefont {Sun}}, \bibinfo
  {author} {\bibfnamefont {F.}~\bibnamefont {Zhou}}, \ and\ \bibinfo {author}
  {\bibfnamefont {Z.-X.}\ \bibnamefont {Zhao}},\ }\href@noop {} {\bibfield
  {journal} {\bibinfo  {journal} {Chin. Phys. Lett.}\ }\textbf {\bibinfo
  {volume} {25}},\ \bibinfo {pages} {2215} (\bibinfo {year}
  {2008})}\BibitemShut {NoStop}%
\bibitem [{\citenamefont {Yang}\ \emph {et~al.}(2008)\citenamefont {Yang},
  \citenamefont {Li}, \citenamefont {Lu}, \citenamefont {Yi}, \citenamefont
  {Shen}, \citenamefont {Ren}, \citenamefont {Che}, \citenamefont {Dong},
  \citenamefont {Sun}, \citenamefont {Zhou},\ and\ \citenamefont
  {Zhao}}]{Yang08SST}%
  \BibitemOpen
  \bibfield  {author} {\bibinfo {author} {\bibfnamefont {J.}~\bibnamefont
  {Yang}}, \bibinfo {author} {\bibfnamefont {Z.-C.}\ \bibnamefont {Li}},
  \bibinfo {author} {\bibfnamefont {W.}~\bibnamefont {Lu}}, \bibinfo {author}
  {\bibfnamefont {W.}~\bibnamefont {Yi}}, \bibinfo {author} {\bibfnamefont
  {X.-L.}\ \bibnamefont {Shen}}, \bibinfo {author} {\bibfnamefont {Z.-A.}\
  \bibnamefont {Ren}}, \bibinfo {author} {\bibfnamefont {G.-C.}\ \bibnamefont
  {Che}}, \bibinfo {author} {\bibfnamefont {X.-L.}\ \bibnamefont {Dong}},
  \bibinfo {author} {\bibfnamefont {L.-L.}\ \bibnamefont {Sun}}, \bibinfo
  {author} {\bibfnamefont {F.}~\bibnamefont {Zhou}}, \ and\ \bibinfo {author}
  {\bibfnamefont {Z.-X.}\ \bibnamefont {Zhao}},\ }\href
  {http://stacks.iop.org/0953-2048/21/082001} {\bibfield  {journal} {\bibinfo
  {journal} {Supercond. Sci. Technol.}\ }\textbf {\bibinfo {volume} {21}},\
  \bibinfo {pages} {082001} (\bibinfo {year} {2008})}\BibitemShut {NoStop}%
\bibitem [{\citenamefont {Wang}\ \emph {et~al.}(2008)\citenamefont {Wang},
  \citenamefont {Li}, \citenamefont {Chi}, \citenamefont {Zhu}, \citenamefont
  {Ren}, \citenamefont {Li}, \citenamefont {Wang}, \citenamefont {Lin},
  \citenamefont {Luo}, \citenamefont {Jiang}, \citenamefont {Xu}, \citenamefont
  {Cao},\ and\ \citenamefont {Xu}}]{Wang08EPL}%
  \BibitemOpen
  \bibfield  {author} {\bibinfo {author} {\bibfnamefont {C.}~\bibnamefont
  {Wang}}, \bibinfo {author} {\bibfnamefont {L.}~\bibnamefont {Li}}, \bibinfo
  {author} {\bibfnamefont {S.}~\bibnamefont {Chi}}, \bibinfo {author}
  {\bibfnamefont {Z.}~\bibnamefont {Zhu}}, \bibinfo {author} {\bibfnamefont
  {Z.}~\bibnamefont {Ren}}, \bibinfo {author} {\bibfnamefont {Y.}~\bibnamefont
  {Li}}, \bibinfo {author} {\bibfnamefont {Y.}~\bibnamefont {Wang}}, \bibinfo
  {author} {\bibfnamefont {X.}~\bibnamefont {Lin}}, \bibinfo {author}
  {\bibfnamefont {Y.}~\bibnamefont {Luo}}, \bibinfo {author} {\bibfnamefont
  {S.}~\bibnamefont {Jiang}}, \bibinfo {author} {\bibfnamefont
  {X.}~\bibnamefont {Xu}}, \bibinfo {author} {\bibfnamefont {G.}~\bibnamefont
  {Cao}}, \ and\ \bibinfo {author} {\bibfnamefont {Z.}~\bibnamefont {Xu}},\
  }\href {http://stacks.iop.org/0295-5075/83/67006} {\bibfield  {journal}
  {\bibinfo  {journal} {Europhys. Lett.}\ }\textbf {\bibinfo {volume} {83}},\
  \bibinfo {pages} {67006} (\bibinfo {year} {2008})}\BibitemShut {NoStop}%
\bibitem [{\citenamefont {Hirschfeld}\ \emph {et~al.}(2011)\citenamefont
  {Hirschfeld}, \citenamefont {Korshunov},\ and\ \citenamefont
  {Mazin}}]{Hirschfeld11RPP}%
  \BibitemOpen
  \bibfield  {author} {\bibinfo {author} {\bibfnamefont {P.~J.}\ \bibnamefont
  {Hirschfeld}}, \bibinfo {author} {\bibfnamefont {M.~M.}\ \bibnamefont
  {Korshunov}}, \ and\ \bibinfo {author} {\bibfnamefont {I.~I.}\ \bibnamefont
  {Mazin}},\ }\href {http://stacks.iop.org/0034-4885/74/i=12/a=124508}
  {\bibfield  {journal} {\bibinfo  {journal} {Rep. Prog. Phys.}\ }\textbf
  {\bibinfo {volume} {74}},\ \bibinfo {pages} {124508} (\bibinfo {year}
  {2011})}\BibitemShut {NoStop}%
\bibitem [{\citenamefont {Mazin}\ \emph {et~al.}(2008)\citenamefont {Mazin},
  \citenamefont {Singh}, \citenamefont {Johannes},\ and\ \citenamefont
  {Du}}]{Mazin08PRL}%
  \BibitemOpen
  \bibfield  {author} {\bibinfo {author} {\bibfnamefont {I.~I.}\ \bibnamefont
  {Mazin}}, \bibinfo {author} {\bibfnamefont {D.~J.}\ \bibnamefont {Singh}},
  \bibinfo {author} {\bibfnamefont {M.~D.}\ \bibnamefont {Johannes}}, \ and\
  \bibinfo {author} {\bibfnamefont {M.~H.}\ \bibnamefont {Du}},\ }\href
  {\doibase 10.1103/PhysRevLett.101.057003} {\bibfield  {journal} {\bibinfo
  {journal} {Phys. Rev. Lett.}\ }\textbf {\bibinfo {volume} {101}},\ \bibinfo
  {pages} {057003} (\bibinfo {year} {2008})}\BibitemShut {NoStop}%
\bibitem [{\citenamefont {Kuroki}\ \emph {et~al.}(2008)\citenamefont {Kuroki},
  \citenamefont {Onari}, \citenamefont {Arita}, \citenamefont {Usui},
  \citenamefont {Tanaka}, \citenamefont {Kontani},\ and\ \citenamefont
  {Aoki}}]{Kuroki08PRL}%
  \BibitemOpen
  \bibfield  {author} {\bibinfo {author} {\bibfnamefont {K.}~\bibnamefont
  {Kuroki}}, \bibinfo {author} {\bibfnamefont {S.}~\bibnamefont {Onari}},
  \bibinfo {author} {\bibfnamefont {R.}~\bibnamefont {Arita}}, \bibinfo
  {author} {\bibfnamefont {H.}~\bibnamefont {Usui}}, \bibinfo {author}
  {\bibfnamefont {Y.}~\bibnamefont {Tanaka}}, \bibinfo {author} {\bibfnamefont
  {H.}~\bibnamefont {Kontani}}, \ and\ \bibinfo {author} {\bibfnamefont
  {H.}~\bibnamefont {Aoki}},\ }\href {\doibase 10.1103/PhysRevLett.101.087004}
  {\bibfield  {journal} {\bibinfo  {journal} {Phys. Rev. Lett.}\ }\textbf
  {\bibinfo {volume} {101}},\ \bibinfo {pages} {087004} (\bibinfo {year}
  {2008})}\BibitemShut {NoStop}%
\bibitem [{\citenamefont {Kontani}\ and\ \citenamefont
  {Onari}(2010)}]{Kontani10PRL}%
  \BibitemOpen
  \bibfield  {author} {\bibinfo {author} {\bibfnamefont {H.}~\bibnamefont
  {Kontani}}\ and\ \bibinfo {author} {\bibfnamefont {S.}~\bibnamefont
  {Onari}},\ }\href {\doibase 10.1103/PhysRevLett.104.157001} {\bibfield
  {journal} {\bibinfo  {journal} {Phys. Rev. Lett.}\ }\textbf {\bibinfo
  {volume} {104}},\ \bibinfo {pages} {157001} (\bibinfo {year}
  {2010})}\BibitemShut {NoStop}%
\bibitem [{\citenamefont {Rotter}\ \emph
  {et~al.}(2008{\natexlab{a}})\citenamefont {Rotter}, \citenamefont {Tegel},\
  and\ \citenamefont {Johrendt}}]{Rotter08PRL}%
  \BibitemOpen
  \bibfield  {author} {\bibinfo {author} {\bibfnamefont {M.}~\bibnamefont
  {Rotter}}, \bibinfo {author} {\bibfnamefont {M.}~\bibnamefont {Tegel}}, \
  and\ \bibinfo {author} {\bibfnamefont {D.}~\bibnamefont {Johrendt}},\ }\href
  {\doibase 10.1103/PhysRevLett.101.107006} {\bibfield  {journal} {\bibinfo
  {journal} {Phys. Rev. Lett.}\ }\textbf {\bibinfo {volume} {101}},\ \bibinfo
  {eid} {107006} (\bibinfo {year} {2008}{\natexlab{a}})}\BibitemShut {NoStop}%
\bibitem [{\citenamefont {Rotter}\ \emph
  {et~al.}(2008{\natexlab{b}})\citenamefont {Rotter}, \citenamefont {Tegel},
  \citenamefont {Johrendt}, \citenamefont {Schellenberg}, \citenamefont
  {Hermes},\ and\ \citenamefont {P\"{o}ttgen}}]{Rotter08PRB}%
  \BibitemOpen
  \bibfield  {author} {\bibinfo {author} {\bibfnamefont {M.}~\bibnamefont
  {Rotter}}, \bibinfo {author} {\bibfnamefont {M.}~\bibnamefont {Tegel}},
  \bibinfo {author} {\bibfnamefont {D.}~\bibnamefont {Johrendt}}, \bibinfo
  {author} {\bibfnamefont {I.}~\bibnamefont {Schellenberg}}, \bibinfo {author}
  {\bibfnamefont {W.}~\bibnamefont {Hermes}}, \ and\ \bibinfo {author}
  {\bibfnamefont {R.}~\bibnamefont {P\"{o}ttgen}},\ }\href {\doibase
  10.1103/PhysRevB.78.020503} {\bibfield  {journal} {\bibinfo  {journal} {Phys.
  Rev. B}\ }\textbf {\bibinfo {volume} {78}},\ \bibinfo {eid} {020503}
  (\bibinfo {year} {2008}{\natexlab{b}})}\BibitemShut {NoStop}%
\bibitem [{\citenamefont {Analytis}\ \emph {et~al.}(2009)\citenamefont
  {Analytis}, \citenamefont {McDonald}, \citenamefont {Chu}, \citenamefont
  {Riggs}, \citenamefont {Bangura}, \citenamefont {Kucharczyk}, \citenamefont
  {Johannes},\ and\ \citenamefont {Fisher}}]{Analytis09PRB}%
  \BibitemOpen
  \bibfield  {author} {\bibinfo {author} {\bibfnamefont {J.~G.}\ \bibnamefont
  {Analytis}}, \bibinfo {author} {\bibfnamefont {R.~D.}\ \bibnamefont
  {McDonald}}, \bibinfo {author} {\bibfnamefont {J.-H.}\ \bibnamefont {Chu}},
  \bibinfo {author} {\bibfnamefont {S.~C.}\ \bibnamefont {Riggs}}, \bibinfo
  {author} {\bibfnamefont {A.~F.}\ \bibnamefont {Bangura}}, \bibinfo {author}
  {\bibfnamefont {C.}~\bibnamefont {Kucharczyk}}, \bibinfo {author}
  {\bibfnamefont {M.}~\bibnamefont {Johannes}}, \ and\ \bibinfo {author}
  {\bibfnamefont {I.~R.}\ \bibnamefont {Fisher}},\ }\href {\doibase
  10.1103/PhysRevB.80.064507} {\bibfield  {journal} {\bibinfo  {journal} {Phys.
  Rev. B}\ }\textbf {\bibinfo {volume} {80}},\ \bibinfo {eid} {064507}
  (\bibinfo {year} {2009})}\BibitemShut {NoStop}%
\bibitem [{\citenamefont {Terashima}\ \emph {et~al.}(2011)\citenamefont
  {Terashima}, \citenamefont {Kurita}, \citenamefont {Tomita}, \citenamefont
  {Kihou}, \citenamefont {Lee}, \citenamefont {Tomioka}, \citenamefont {Ito},
  \citenamefont {Iyo}, \citenamefont {Eisaki}, \citenamefont {Liang},
  \citenamefont {Nakajima}, \citenamefont {Ishida}, \citenamefont {Uchida},
  \citenamefont {Harima},\ and\ \citenamefont {Uji}}]{Terashima11PRL}%
  \BibitemOpen
  \bibfield  {author} {\bibinfo {author} {\bibfnamefont {T.}~\bibnamefont
  {Terashima}}, \bibinfo {author} {\bibfnamefont {N.}~\bibnamefont {Kurita}},
  \bibinfo {author} {\bibfnamefont {M.}~\bibnamefont {Tomita}}, \bibinfo
  {author} {\bibfnamefont {K.}~\bibnamefont {Kihou}}, \bibinfo {author}
  {\bibfnamefont {C.~H.}\ \bibnamefont {Lee}}, \bibinfo {author} {\bibfnamefont
  {Y.}~\bibnamefont {Tomioka}}, \bibinfo {author} {\bibfnamefont
  {T.}~\bibnamefont {Ito}}, \bibinfo {author} {\bibfnamefont {A.}~\bibnamefont
  {Iyo}}, \bibinfo {author} {\bibfnamefont {H.}~\bibnamefont {Eisaki}},
  \bibinfo {author} {\bibfnamefont {T.}~\bibnamefont {Liang}}, \bibinfo
  {author} {\bibfnamefont {M.}~\bibnamefont {Nakajima}}, \bibinfo {author}
  {\bibfnamefont {S.}~\bibnamefont {Ishida}}, \bibinfo {author} {\bibfnamefont
  {S.}~\bibnamefont {Uchida}}, \bibinfo {author} {\bibfnamefont
  {H.}~\bibnamefont {Harima}}, \ and\ \bibinfo {author} {\bibfnamefont
  {S.}~\bibnamefont {Uji}},\ }\href {\doibase 10.1103/PhysRevLett.107.176402}
  {\bibfield  {journal} {\bibinfo  {journal} {Phys. Rev. Lett.}\ }\textbf
  {\bibinfo {volume} {107}},\ \bibinfo {pages} {176402} (\bibinfo {year}
  {2011})}\BibitemShut {NoStop}%
\bibitem [{\citenamefont {Rotter}\ \emph
  {et~al.}(2008{\natexlab{c}})\citenamefont {Rotter}, \citenamefont {Pangerl},
  \citenamefont {Tegel},\ and\ \citenamefont {Johrendt}}]{Rotter08ACIE}%
  \BibitemOpen
  \bibfield  {author} {\bibinfo {author} {\bibfnamefont {M.}~\bibnamefont
  {Rotter}}, \bibinfo {author} {\bibfnamefont {M.}~\bibnamefont {Pangerl}},
  \bibinfo {author} {\bibfnamefont {M.}~\bibnamefont {Tegel}}, \ and\ \bibinfo
  {author} {\bibfnamefont {D.}~\bibnamefont {Johrendt}},\ }\href@noop {}
  {\bibfield  {journal} {\bibinfo  {journal} {Angew. Chem. Int. Ed.}\ }\textbf
  {\bibinfo {volume} {47}},\ \bibinfo {pages} {7949} (\bibinfo {year}
  {2008}{\natexlab{c}})}\BibitemShut {NoStop}%
\bibitem [{\citenamefont {Sasmal}\ \emph {et~al.}(2008)\citenamefont {Sasmal},
  \citenamefont {Lv}, \citenamefont {Lorenz}, \citenamefont {Guloy},
  \citenamefont {Chen}, \citenamefont {Xue},\ and\ \citenamefont
  {Chu}}]{Sasmal08PRL}%
  \BibitemOpen
  \bibfield  {author} {\bibinfo {author} {\bibfnamefont {K.}~\bibnamefont
  {Sasmal}}, \bibinfo {author} {\bibfnamefont {B.}~\bibnamefont {Lv}}, \bibinfo
  {author} {\bibfnamefont {B.}~\bibnamefont {Lorenz}}, \bibinfo {author}
  {\bibfnamefont {A.~M.}\ \bibnamefont {Guloy}}, \bibinfo {author}
  {\bibfnamefont {F.}~\bibnamefont {Chen}}, \bibinfo {author} {\bibfnamefont
  {Y.-Y.}\ \bibnamefont {Xue}}, \ and\ \bibinfo {author} {\bibfnamefont
  {C.-W.}\ \bibnamefont {Chu}},\ }\href {\doibase
  10.1103/PhysRevLett.101.107007} {\bibfield  {journal} {\bibinfo  {journal}
  {Phys. Rev. Lett.}\ }\textbf {\bibinfo {volume} {101}},\ \bibinfo {eid}
  {107007} (\bibinfo {year} {2008})}\BibitemShut {NoStop}%
\bibitem [{\citenamefont {Chen}\ \emph {et~al.}(2009)\citenamefont {Chen},
  \citenamefont {Ren}, \citenamefont {Qiu}, \citenamefont {Bao}, \citenamefont
  {Liu}, \citenamefont {Wu}, \citenamefont {Wu}, \citenamefont {Xie},
  \citenamefont {Wang}, \citenamefont {Huang},\ and\ \citenamefont
  {Chen}}]{Chen09EPL}%
  \BibitemOpen
  \bibfield  {author} {\bibinfo {author} {\bibfnamefont {H.}~\bibnamefont
  {Chen}}, \bibinfo {author} {\bibfnamefont {Y.}~\bibnamefont {Ren}}, \bibinfo
  {author} {\bibfnamefont {Y.}~\bibnamefont {Qiu}}, \bibinfo {author}
  {\bibfnamefont {W.}~\bibnamefont {Bao}}, \bibinfo {author} {\bibfnamefont
  {R.~H.}\ \bibnamefont {Liu}}, \bibinfo {author} {\bibfnamefont
  {G.}~\bibnamefont {Wu}}, \bibinfo {author} {\bibfnamefont {T.}~\bibnamefont
  {Wu}}, \bibinfo {author} {\bibfnamefont {Y.~L.}\ \bibnamefont {Xie}},
  \bibinfo {author} {\bibfnamefont {X.~F.}\ \bibnamefont {Wang}}, \bibinfo
  {author} {\bibfnamefont {Q.}~\bibnamefont {Huang}}, \ and\ \bibinfo {author}
  {\bibfnamefont {X.~H.}\ \bibnamefont {Chen}},\ }\href
  {http://stacks.iop.org/0295-5075/85/17006} {\bibfield  {journal} {\bibinfo
  {journal} {Europhys. Lett.}\ }\textbf {\bibinfo {volume} {85}},\ \bibinfo
  {pages} {17006} (\bibinfo {year} {2009})}\BibitemShut {NoStop}%
\bibitem [{\citenamefont {Kihou}\ \emph {et~al.}(2010)\citenamefont {Kihou},
  \citenamefont {Saito}, \citenamefont {Ishida}, \citenamefont {Nakajima},
  \citenamefont {Tomioka}, \citenamefont {Fukazawa}, \citenamefont {Kohori},
  \citenamefont {Ito}, \citenamefont {Uchida}, \citenamefont {Iyo},
  \citenamefont {Lee},\ and\ \citenamefont {Eisaki}}]{Kihou10JPSJ}%
  \BibitemOpen
  \bibfield  {author} {\bibinfo {author} {\bibfnamefont {K.}~\bibnamefont
  {Kihou}}, \bibinfo {author} {\bibfnamefont {T.}~\bibnamefont {Saito}},
  \bibinfo {author} {\bibfnamefont {S.}~\bibnamefont {Ishida}}, \bibinfo
  {author} {\bibfnamefont {M.}~\bibnamefont {Nakajima}}, \bibinfo {author}
  {\bibfnamefont {Y.}~\bibnamefont {Tomioka}}, \bibinfo {author} {\bibfnamefont
  {H.}~\bibnamefont {Fukazawa}}, \bibinfo {author} {\bibfnamefont
  {Y.}~\bibnamefont {Kohori}}, \bibinfo {author} {\bibfnamefont
  {T.}~\bibnamefont {Ito}}, \bibinfo {author} {\bibfnamefont {S.}~\bibnamefont
  {Uchida}}, \bibinfo {author} {\bibfnamefont {A.}~\bibnamefont {Iyo}},
  \bibinfo {author} {\bibfnamefont {C.~H.}\ \bibnamefont {Lee}}, \ and\
  \bibinfo {author} {\bibfnamefont {H.}~\bibnamefont {Eisaki}},\ }\href
  {\doibase 10.1143/JPSJ.79.124713} {\bibfield  {journal} {\bibinfo  {journal}
  {J. Phys. Soc. Jpn.}\ }\textbf {\bibinfo {volume} {79}},\ \bibinfo {pages}
  {124713} (\bibinfo {year} {2010})}\BibitemShut {NoStop}%
\bibitem [{\citenamefont {Hashimoto}\ \emph {et~al.}(2009)\citenamefont
  {Hashimoto}, \citenamefont {Shibauchi}, \citenamefont {Kasahara},
  \citenamefont {Ikada}, \citenamefont {Tonegawa}, \citenamefont {Kato},
  \citenamefont {Okazaki}, \citenamefont {van~der Beek}, \citenamefont
  {Konczykowski}, \citenamefont {Takeya}, \citenamefont {Hirata}, \citenamefont
  {Terashima},\ and\ \citenamefont {Matsuda}}]{Hashimoto09PRL}%
  \BibitemOpen
  \bibfield  {author} {\bibinfo {author} {\bibfnamefont {K.}~\bibnamefont
  {Hashimoto}}, \bibinfo {author} {\bibfnamefont {T.}~\bibnamefont
  {Shibauchi}}, \bibinfo {author} {\bibfnamefont {S.}~\bibnamefont {Kasahara}},
  \bibinfo {author} {\bibfnamefont {K.}~\bibnamefont {Ikada}}, \bibinfo
  {author} {\bibfnamefont {S.}~\bibnamefont {Tonegawa}}, \bibinfo {author}
  {\bibfnamefont {T.}~\bibnamefont {Kato}}, \bibinfo {author} {\bibfnamefont
  {R.}~\bibnamefont {Okazaki}}, \bibinfo {author} {\bibfnamefont {C.~J.}\
  \bibnamefont {van~der Beek}}, \bibinfo {author} {\bibfnamefont
  {M.}~\bibnamefont {Konczykowski}}, \bibinfo {author} {\bibfnamefont
  {H.}~\bibnamefont {Takeya}}, \bibinfo {author} {\bibfnamefont
  {K.}~\bibnamefont {Hirata}}, \bibinfo {author} {\bibfnamefont
  {T.}~\bibnamefont {Terashima}}, \ and\ \bibinfo {author} {\bibfnamefont
  {Y.}~\bibnamefont {Matsuda}},\ }\href {\doibase
  10.1103/PhysRevLett.102.207001} {\bibfield  {journal} {\bibinfo  {journal}
  {Phys. Rev. Lett.}\ }\textbf {\bibinfo {volume} {102}},\ \bibinfo {pages}
  {207001} (\bibinfo {year} {2009})}\BibitemShut {NoStop}%
\bibitem [{\citenamefont {Mu}\ \emph {et~al.}(2009)\citenamefont {Mu},
  \citenamefont {Luo}, \citenamefont {Wang}, \citenamefont {Shan},
  \citenamefont {Ren},\ and\ \citenamefont {Wen}}]{Mu09PRB}%
  \BibitemOpen
  \bibfield  {author} {\bibinfo {author} {\bibfnamefont {G.}~\bibnamefont
  {Mu}}, \bibinfo {author} {\bibfnamefont {H.}~\bibnamefont {Luo}}, \bibinfo
  {author} {\bibfnamefont {Z.}~\bibnamefont {Wang}}, \bibinfo {author}
  {\bibfnamefont {L.}~\bibnamefont {Shan}}, \bibinfo {author} {\bibfnamefont
  {C.}~\bibnamefont {Ren}}, \ and\ \bibinfo {author} {\bibfnamefont {H.-H.}\
  \bibnamefont {Wen}},\ }\href {\doibase 10.1103/PhysRevB.79.174501} {\bibfield
   {journal} {\bibinfo  {journal} {Phys. Rev. B}\ }\textbf {\bibinfo {volume}
  {79}},\ \bibinfo {pages} {174501} (\bibinfo {year} {2009})}\BibitemShut
  {NoStop}%
\bibitem [{\citenamefont {Popovich}\ \emph {et~al.}(2010)\citenamefont
  {Popovich}, \citenamefont {Boris}, \citenamefont {Dolgov}, \citenamefont
  {Golubov}, \citenamefont {Sun}, \citenamefont {Lin}, \citenamefont {Kremer},\
  and\ \citenamefont {Keimer}}]{Popovich10PRL}%
  \BibitemOpen
  \bibfield  {author} {\bibinfo {author} {\bibfnamefont {P.}~\bibnamefont
  {Popovich}}, \bibinfo {author} {\bibfnamefont {A.~V.}\ \bibnamefont {Boris}},
  \bibinfo {author} {\bibfnamefont {O.~V.}\ \bibnamefont {Dolgov}}, \bibinfo
  {author} {\bibfnamefont {A.~A.}\ \bibnamefont {Golubov}}, \bibinfo {author}
  {\bibfnamefont {D.~L.}\ \bibnamefont {Sun}}, \bibinfo {author} {\bibfnamefont
  {C.~T.}\ \bibnamefont {Lin}}, \bibinfo {author} {\bibfnamefont {R.~K.}\
  \bibnamefont {Kremer}}, \ and\ \bibinfo {author} {\bibfnamefont
  {B.}~\bibnamefont {Keimer}},\ }\href {\doibase
  10.1103/PhysRevLett.105.027003} {\bibfield  {journal} {\bibinfo  {journal}
  {Phys. Rev. Lett.}\ }\textbf {\bibinfo {volume} {105}},\ \bibinfo {pages}
  {027003} (\bibinfo {year} {2010})}\BibitemShut {NoStop}%
\bibitem [{\citenamefont {Luo}\ \emph {et~al.}(2009)\citenamefont {Luo},
  \citenamefont {Tanatar}, \citenamefont {Reid}, \citenamefont {Shakeripour},
  \citenamefont {Doiron-Leyraud}, \citenamefont {Ni}, \citenamefont {Bud'ko},
  \citenamefont {Canfield}, \citenamefont {Luo}, \citenamefont {Wang},
  \citenamefont {Wen}, \citenamefont {Prozorov},\ and\ \citenamefont
  {Taillefer}}]{Luo09PRB}%
  \BibitemOpen
  \bibfield  {author} {\bibinfo {author} {\bibfnamefont {X.~G.}\ \bibnamefont
  {Luo}}, \bibinfo {author} {\bibfnamefont {M.~A.}\ \bibnamefont {Tanatar}},
  \bibinfo {author} {\bibfnamefont {J.-P.}\ \bibnamefont {Reid}}, \bibinfo
  {author} {\bibfnamefont {H.}~\bibnamefont {Shakeripour}}, \bibinfo {author}
  {\bibfnamefont {N.}~\bibnamefont {Doiron-Leyraud}}, \bibinfo {author}
  {\bibfnamefont {N.}~\bibnamefont {Ni}}, \bibinfo {author} {\bibfnamefont
  {S.~L.}\ \bibnamefont {Bud'ko}}, \bibinfo {author} {\bibfnamefont {P.~C.}\
  \bibnamefont {Canfield}}, \bibinfo {author} {\bibfnamefont {H.}~\bibnamefont
  {Luo}}, \bibinfo {author} {\bibfnamefont {Z.}~\bibnamefont {Wang}}, \bibinfo
  {author} {\bibfnamefont {H.-H.}\ \bibnamefont {Wen}}, \bibinfo {author}
  {\bibfnamefont {R.}~\bibnamefont {Prozorov}}, \ and\ \bibinfo {author}
  {\bibfnamefont {L.}~\bibnamefont {Taillefer}},\ }\href {\doibase
  10.1103/PhysRevB.80.140503} {\bibfield  {journal} {\bibinfo  {journal} {Phys.
  Rev. B}\ }\textbf {\bibinfo {volume} {80}},\ \bibinfo {pages} {140503}
  (\bibinfo {year} {2009})}\BibitemShut {NoStop}%
\bibitem [{\citenamefont {Yashima}\ \emph {et~al.}(2009)\citenamefont
  {Yashima}, \citenamefont {Nishimura}, \citenamefont {Mukuda}, \citenamefont
  {Kitaoka}, \citenamefont {Miyazawa}, \citenamefont {Shirage}, \citenamefont
  {Kihou}, \citenamefont {Kito}, \citenamefont {Eisaki},\ and\ \citenamefont
  {Iyo}}]{Yashima09JPSJ}%
  \BibitemOpen
  \bibfield  {author} {\bibinfo {author} {\bibfnamefont {M.}~\bibnamefont
  {Yashima}}, \bibinfo {author} {\bibfnamefont {H.}~\bibnamefont {Nishimura}},
  \bibinfo {author} {\bibfnamefont {H.}~\bibnamefont {Mukuda}}, \bibinfo
  {author} {\bibfnamefont {Y.}~\bibnamefont {Kitaoka}}, \bibinfo {author}
  {\bibfnamefont {K.}~\bibnamefont {Miyazawa}}, \bibinfo {author}
  {\bibfnamefont {P.~M.}\ \bibnamefont {Shirage}}, \bibinfo {author}
  {\bibfnamefont {K.}~\bibnamefont {Kihou}}, \bibinfo {author} {\bibfnamefont
  {H.}~\bibnamefont {Kito}}, \bibinfo {author} {\bibfnamefont {H.}~\bibnamefont
  {Eisaki}}, \ and\ \bibinfo {author} {\bibfnamefont {A.}~\bibnamefont {Iyo}},\
  }\href {\doibase 10.1143/JPSJ.78.103702} {\bibfield  {journal} {\bibinfo
  {journal} {J. Phys. Soc. Jpn.}\ }\textbf {\bibinfo {volume} {78}},\ \bibinfo
  {pages} {103702} (\bibinfo {year} {2009})}\BibitemShut {NoStop}%
\bibitem [{\citenamefont {Fukazawa}\ \emph {et~al.}(2009)\citenamefont
  {Fukazawa}, \citenamefont {Yamada}, \citenamefont {Kondo}, \citenamefont
  {Saito}, \citenamefont {Kohori}, \citenamefont {Kuga}, \citenamefont
  {Matsumoto}, \citenamefont {Nakatsuji}, \citenamefont {Kito}, \citenamefont
  {Shirage}, \citenamefont {Kihou}, \citenamefont {Takeshita}, \citenamefont
  {Lee}, \citenamefont {Iyo},\ and\ \citenamefont
  {Eisaki}}]{Fukazawa09JPSJ_KFA}%
  \BibitemOpen
  \bibfield  {author} {\bibinfo {author} {\bibfnamefont {H.}~\bibnamefont
  {Fukazawa}}, \bibinfo {author} {\bibfnamefont {Y.}~\bibnamefont {Yamada}},
  \bibinfo {author} {\bibfnamefont {K.}~\bibnamefont {Kondo}}, \bibinfo
  {author} {\bibfnamefont {T.}~\bibnamefont {Saito}}, \bibinfo {author}
  {\bibfnamefont {Y.}~\bibnamefont {Kohori}}, \bibinfo {author} {\bibfnamefont
  {K.}~\bibnamefont {Kuga}}, \bibinfo {author} {\bibfnamefont {Y.}~\bibnamefont
  {Matsumoto}}, \bibinfo {author} {\bibfnamefont {S.}~\bibnamefont
  {Nakatsuji}}, \bibinfo {author} {\bibfnamefont {H.}~\bibnamefont {Kito}},
  \bibinfo {author} {\bibfnamefont {P.~M.}\ \bibnamefont {Shirage}}, \bibinfo
  {author} {\bibfnamefont {K.}~\bibnamefont {Kihou}}, \bibinfo {author}
  {\bibfnamefont {N.}~\bibnamefont {Takeshita}}, \bibinfo {author}
  {\bibfnamefont {C.~H.}\ \bibnamefont {Lee}}, \bibinfo {author} {\bibfnamefont
  {A.}~\bibnamefont {Iyo}}, \ and\ \bibinfo {author} {\bibfnamefont
  {H.}~\bibnamefont {Eisaki}},\ }\href {\doibase 10.1143/JPSJ.78.083712}
  {\bibfield  {journal} {\bibinfo  {journal} {J. Phys. Soc. Jpn.}\ }\textbf
  {\bibinfo {volume} {78}},\ \bibinfo {pages} {083712} (\bibinfo {year}
  {2009})}\BibitemShut {NoStop}%
\bibitem [{\citenamefont {Hashimoto}\ \emph {et~al.}(2010)\citenamefont
  {Hashimoto}, \citenamefont {Serafin}, \citenamefont {Tonegawa}, \citenamefont
  {Katsumata}, \citenamefont {Okazaki}, \citenamefont {Saito}, \citenamefont
  {Fukazawa}, \citenamefont {Kohori}, \citenamefont {Kihou}, \citenamefont
  {Lee}, \citenamefont {Iyo}, \citenamefont {Eisaki}, \citenamefont {Ikeda},
  \citenamefont {Matsuda}, \citenamefont {Carrington},\ and\ \citenamefont
  {Shibauchi}}]{Hashimoto10PRB}%
  \BibitemOpen
  \bibfield  {author} {\bibinfo {author} {\bibfnamefont {K.}~\bibnamefont
  {Hashimoto}}, \bibinfo {author} {\bibfnamefont {A.}~\bibnamefont {Serafin}},
  \bibinfo {author} {\bibfnamefont {S.}~\bibnamefont {Tonegawa}}, \bibinfo
  {author} {\bibfnamefont {R.}~\bibnamefont {Katsumata}}, \bibinfo {author}
  {\bibfnamefont {R.}~\bibnamefont {Okazaki}}, \bibinfo {author} {\bibfnamefont
  {T.}~\bibnamefont {Saito}}, \bibinfo {author} {\bibfnamefont
  {H.}~\bibnamefont {Fukazawa}}, \bibinfo {author} {\bibfnamefont
  {Y.}~\bibnamefont {Kohori}}, \bibinfo {author} {\bibfnamefont
  {K.}~\bibnamefont {Kihou}}, \bibinfo {author} {\bibfnamefont {C.~H.}\
  \bibnamefont {Lee}}, \bibinfo {author} {\bibfnamefont {A.}~\bibnamefont
  {Iyo}}, \bibinfo {author} {\bibfnamefont {H.}~\bibnamefont {Eisaki}},
  \bibinfo {author} {\bibfnamefont {H.}~\bibnamefont {Ikeda}}, \bibinfo
  {author} {\bibfnamefont {Y.}~\bibnamefont {Matsuda}}, \bibinfo {author}
  {\bibfnamefont {A.}~\bibnamefont {Carrington}}, \ and\ \bibinfo {author}
  {\bibfnamefont {T.}~\bibnamefont {Shibauchi}},\ }\href {\doibase
  10.1103/PhysRevB.82.014526} {\bibfield  {journal} {\bibinfo  {journal} {Phys.
  Rev. B}\ }\textbf {\bibinfo {volume} {82}},\ \bibinfo {pages} {014526}
  (\bibinfo {year} {2010})}\BibitemShut {NoStop}%
\bibitem [{\citenamefont {Dong}\ \emph {et~al.}(2010)\citenamefont {Dong},
  \citenamefont {Zhou}, \citenamefont {Guan}, \citenamefont {Zhang},
  \citenamefont {Dai}, \citenamefont {Qiu}, \citenamefont {Wang}, \citenamefont
  {He}, \citenamefont {Chen},\ and\ \citenamefont {Li}}]{Dong10PRL}%
  \BibitemOpen
  \bibfield  {author} {\bibinfo {author} {\bibfnamefont {J.~K.}\ \bibnamefont
  {Dong}}, \bibinfo {author} {\bibfnamefont {S.~Y.}\ \bibnamefont {Zhou}},
  \bibinfo {author} {\bibfnamefont {T.~Y.}\ \bibnamefont {Guan}}, \bibinfo
  {author} {\bibfnamefont {H.}~\bibnamefont {Zhang}}, \bibinfo {author}
  {\bibfnamefont {Y.~F.}\ \bibnamefont {Dai}}, \bibinfo {author} {\bibfnamefont
  {X.}~\bibnamefont {Qiu}}, \bibinfo {author} {\bibfnamefont {X.~F.}\
  \bibnamefont {Wang}}, \bibinfo {author} {\bibfnamefont {Y.}~\bibnamefont
  {He}}, \bibinfo {author} {\bibfnamefont {X.~H.}\ \bibnamefont {Chen}}, \ and\
  \bibinfo {author} {\bibfnamefont {S.~Y.}\ \bibnamefont {Li}},\ }\href
  {\doibase 10.1103/PhysRevLett.104.087005} {\bibfield  {journal} {\bibinfo
  {journal} {Phys. Rev. Lett.}\ }\textbf {\bibinfo {volume} {104}},\ \bibinfo
  {pages} {087005} (\bibinfo {year} {2010})}\BibitemShut {NoStop}%
\bibitem [{\citenamefont {Terashima}\ \emph
  {et~al.}(2010{\natexlab{a}})\citenamefont {Terashima}, \citenamefont
  {Kimata}, \citenamefont {Kurita}, \citenamefont {Satsukawa}, \citenamefont
  {Harada}, \citenamefont {Hazama}, \citenamefont {Imai}, \citenamefont {Sato},
  \citenamefont {Kihou}, \citenamefont {Lee}, \citenamefont {Kito},
  \citenamefont {Eisaki}, \citenamefont {Iyo}, \citenamefont {Saito},
  \citenamefont {Fukazawa}, \citenamefont {Kohori}, \citenamefont {Harima},\
  and\ \citenamefont {Uji}}]{Terashima10PRL_comment}%
  \BibitemOpen
  \bibfield  {author} {\bibinfo {author} {\bibfnamefont {T.}~\bibnamefont
  {Terashima}}, \bibinfo {author} {\bibfnamefont {M.}~\bibnamefont {Kimata}},
  \bibinfo {author} {\bibfnamefont {N.}~\bibnamefont {Kurita}}, \bibinfo
  {author} {\bibfnamefont {H.}~\bibnamefont {Satsukawa}}, \bibinfo {author}
  {\bibfnamefont {A.}~\bibnamefont {Harada}}, \bibinfo {author} {\bibfnamefont
  {K.}~\bibnamefont {Hazama}}, \bibinfo {author} {\bibfnamefont
  {M.}~\bibnamefont {Imai}}, \bibinfo {author} {\bibfnamefont {A.}~\bibnamefont
  {Sato}}, \bibinfo {author} {\bibfnamefont {K.}~\bibnamefont {Kihou}},
  \bibinfo {author} {\bibfnamefont {C.~H.}\ \bibnamefont {Lee}}, \bibinfo
  {author} {\bibfnamefont {H.}~\bibnamefont {Kito}}, \bibinfo {author}
  {\bibfnamefont {H.}~\bibnamefont {Eisaki}}, \bibinfo {author} {\bibfnamefont
  {A.}~\bibnamefont {Iyo}}, \bibinfo {author} {\bibfnamefont {T.}~\bibnamefont
  {Saito}}, \bibinfo {author} {\bibfnamefont {H.}~\bibnamefont {Fukazawa}},
  \bibinfo {author} {\bibfnamefont {Y.}~\bibnamefont {Kohori}}, \bibinfo
  {author} {\bibfnamefont {H.}~\bibnamefont {Harima}}, \ and\ \bibinfo {author}
  {\bibfnamefont {S.}~\bibnamefont {Uji}},\ }\href {\doibase
  10.1103/PhysRevLett.104.259701} {\bibfield  {journal} {\bibinfo  {journal}
  {Phys. Rev. Lett.}\ }\textbf {\bibinfo {volume} {104}},\ \bibinfo {pages}
  {259701} (\bibinfo {year} {2010}{\natexlab{a}})}\BibitemShut {NoStop}%
\bibitem [{\citenamefont {Reid}\ \emph {et~al.}(2012)\citenamefont {Reid},
  \citenamefont {Tanatar}, \citenamefont {Juneau-Fecteau}, \citenamefont
  {Gordon}, \citenamefont {de~Cotret}, \citenamefont {Doiron-Leyraud},
  \citenamefont {Saito}, \citenamefont {Fukazawa}, \citenamefont {Kohori},
  \citenamefont {Kihou}, \citenamefont {Lee}, \citenamefont {Iyo},
  \citenamefont {Eisaki}, \citenamefont {Prozorov},\ and\ \citenamefont
  {Taillefer}}]{Reid12PRL}%
  \BibitemOpen
  \bibfield  {author} {\bibinfo {author} {\bibfnamefont {J.-P.}\ \bibnamefont
  {Reid}}, \bibinfo {author} {\bibfnamefont {M.~A.}\ \bibnamefont {Tanatar}},
  \bibinfo {author} {\bibfnamefont {A.}~\bibnamefont {Juneau-Fecteau}},
  \bibinfo {author} {\bibfnamefont {R.~T.}\ \bibnamefont {Gordon}}, \bibinfo
  {author} {\bibfnamefont {S.~R.}\ \bibnamefont {de~Cotret}}, \bibinfo {author}
  {\bibfnamefont {N.}~\bibnamefont {Doiron-Leyraud}}, \bibinfo {author}
  {\bibfnamefont {T.}~\bibnamefont {Saito}}, \bibinfo {author} {\bibfnamefont
  {H.}~\bibnamefont {Fukazawa}}, \bibinfo {author} {\bibfnamefont
  {Y.}~\bibnamefont {Kohori}}, \bibinfo {author} {\bibfnamefont
  {K.}~\bibnamefont {Kihou}}, \bibinfo {author} {\bibfnamefont {C.~H.}\
  \bibnamefont {Lee}}, \bibinfo {author} {\bibfnamefont {A.}~\bibnamefont
  {Iyo}}, \bibinfo {author} {\bibfnamefont {H.}~\bibnamefont {Eisaki}},
  \bibinfo {author} {\bibfnamefont {R.}~\bibnamefont {Prozorov}}, \ and\
  \bibinfo {author} {\bibfnamefont {L.}~\bibnamefont {Taillefer}},\ }\href
  {\doibase 10.1103/PhysRevLett.109.087001} {\bibfield  {journal} {\bibinfo
  {journal} {Phys. Rev. Lett.}\ }\textbf {\bibinfo {volume} {109}},\ \bibinfo
  {pages} {087001} (\bibinfo {year} {2012})}\BibitemShut {NoStop}%
\bibitem [{\citenamefont {Malaeb}\ \emph {et~al.}(2012)\citenamefont {Malaeb},
  \citenamefont {Shimojima}, \citenamefont {Ishida}, \citenamefont {Okazaki},
  \citenamefont {Ota}, \citenamefont {Ohgushi}, \citenamefont {Kihou},
  \citenamefont {Saito}, \citenamefont {Lee}, \citenamefont {Ishida},
  \citenamefont {Nakajima}, \citenamefont {Uchida}, \citenamefont {Fukazawa},
  \citenamefont {Kohori}, \citenamefont {Iyo}, \citenamefont {Eisaki},
  \citenamefont {Chen}, \citenamefont {Watanabe}, \citenamefont {Ikeda},\ and\
  \citenamefont {Shin}}]{Malaeb12PRB}%
  \BibitemOpen
  \bibfield  {author} {\bibinfo {author} {\bibfnamefont {W.}~\bibnamefont
  {Malaeb}}, \bibinfo {author} {\bibfnamefont {T.}~\bibnamefont {Shimojima}},
  \bibinfo {author} {\bibfnamefont {Y.}~\bibnamefont {Ishida}}, \bibinfo
  {author} {\bibfnamefont {K.}~\bibnamefont {Okazaki}}, \bibinfo {author}
  {\bibfnamefont {Y.}~\bibnamefont {Ota}}, \bibinfo {author} {\bibfnamefont
  {K.}~\bibnamefont {Ohgushi}}, \bibinfo {author} {\bibfnamefont
  {K.}~\bibnamefont {Kihou}}, \bibinfo {author} {\bibfnamefont
  {T.}~\bibnamefont {Saito}}, \bibinfo {author} {\bibfnamefont {C.~H.}\
  \bibnamefont {Lee}}, \bibinfo {author} {\bibfnamefont {S.}~\bibnamefont
  {Ishida}}, \bibinfo {author} {\bibfnamefont {M.}~\bibnamefont {Nakajima}},
  \bibinfo {author} {\bibfnamefont {S.}~\bibnamefont {Uchida}}, \bibinfo
  {author} {\bibfnamefont {H.}~\bibnamefont {Fukazawa}}, \bibinfo {author}
  {\bibfnamefont {Y.}~\bibnamefont {Kohori}}, \bibinfo {author} {\bibfnamefont
  {A.}~\bibnamefont {Iyo}}, \bibinfo {author} {\bibfnamefont {H.}~\bibnamefont
  {Eisaki}}, \bibinfo {author} {\bibfnamefont {C.-T.}\ \bibnamefont {Chen}},
  \bibinfo {author} {\bibfnamefont {S.}~\bibnamefont {Watanabe}}, \bibinfo
  {author} {\bibfnamefont {H.}~\bibnamefont {Ikeda}}, \ and\ \bibinfo {author}
  {\bibfnamefont {S.}~\bibnamefont {Shin}},\ }\href {\doibase
  10.1103/PhysRevB.86.165117} {\bibfield  {journal} {\bibinfo  {journal} {Phys.
  Rev. B}\ }\textbf {\bibinfo {volume} {86}},\ \bibinfo {pages} {165117}
  (\bibinfo {year} {2012})}\BibitemShut {NoStop}%
\bibitem [{\citenamefont {Hirano}\ \emph {et~al.}(2012)\citenamefont {Hirano},
  \citenamefont {Yamada}, \citenamefont {Saito}, \citenamefont {Nagashima},
  \citenamefont {Konishi}, \citenamefont {Toriyama}, \citenamefont {Ohta},
  \citenamefont {Fukazawa}, \citenamefont {Kohori}, \citenamefont {Furukawa},
  \citenamefont {Kihou}, \citenamefont {Lee}, \citenamefont {Iyo},\ and\
  \citenamefont {Eisaki}}]{Hirano12JPSJ}%
  \BibitemOpen
  \bibfield  {author} {\bibinfo {author} {\bibfnamefont {M.}~\bibnamefont
  {Hirano}}, \bibinfo {author} {\bibfnamefont {Y.}~\bibnamefont {Yamada}},
  \bibinfo {author} {\bibfnamefont {T.}~\bibnamefont {Saito}}, \bibinfo
  {author} {\bibfnamefont {R.}~\bibnamefont {Nagashima}}, \bibinfo {author}
  {\bibfnamefont {T.}~\bibnamefont {Konishi}}, \bibinfo {author} {\bibfnamefont
  {T.}~\bibnamefont {Toriyama}}, \bibinfo {author} {\bibfnamefont
  {Y.}~\bibnamefont {Ohta}}, \bibinfo {author} {\bibfnamefont {H.}~\bibnamefont
  {Fukazawa}}, \bibinfo {author} {\bibfnamefont {Y.}~\bibnamefont {Kohori}},
  \bibinfo {author} {\bibfnamefont {Y.}~\bibnamefont {Furukawa}}, \bibinfo
  {author} {\bibfnamefont {K.}~\bibnamefont {Kihou}}, \bibinfo {author}
  {\bibfnamefont {C.~H.}\ \bibnamefont {Lee}}, \bibinfo {author} {\bibfnamefont
  {A.}~\bibnamefont {Iyo}}, \ and\ \bibinfo {author} {\bibfnamefont
  {H.}~\bibnamefont {Eisaki}},\ }\href {\doibase 10.1143/JPSJ.81.054704}
  {\bibfield  {journal} {\bibinfo  {journal} {J. Phys. Soc. Jpn.}\ }\textbf
  {\bibinfo {volume} {81}},\ \bibinfo {pages} {054704} (\bibinfo {year}
  {2012})}\BibitemShut {NoStop}%
\bibitem [{\citenamefont {Okazaki}\ \emph {et~al.}(2012)\citenamefont
  {Okazaki}, \citenamefont {Ota}, \citenamefont {Kotani}, \citenamefont
  {Malaeb}, \citenamefont {Ishida}, \citenamefont {Shimojima}, \citenamefont
  {Kiss}, \citenamefont {Watanabe}, \citenamefont {Chen}, \citenamefont
  {Kihou}, \citenamefont {Lee}, \citenamefont {Iyo}, \citenamefont {Eisaki},
  \citenamefont {Saito}, \citenamefont {Fukazawa}, \citenamefont {Kohori},
  \citenamefont {Hashimoto}, \citenamefont {Shibauchi}, \citenamefont
  {Matsuda}, \citenamefont {Ikeda}, \citenamefont {Miyahara}, \citenamefont
  {Arita}, \citenamefont {Chainani},\ and\ \citenamefont
  {Shin}}]{Okazaki12Science}%
  \BibitemOpen
  \bibfield  {author} {\bibinfo {author} {\bibfnamefont {K.}~\bibnamefont
  {Okazaki}}, \bibinfo {author} {\bibfnamefont {Y.}~\bibnamefont {Ota}},
  \bibinfo {author} {\bibfnamefont {Y.}~\bibnamefont {Kotani}}, \bibinfo
  {author} {\bibfnamefont {W.}~\bibnamefont {Malaeb}}, \bibinfo {author}
  {\bibfnamefont {Y.}~\bibnamefont {Ishida}}, \bibinfo {author} {\bibfnamefont
  {T.}~\bibnamefont {Shimojima}}, \bibinfo {author} {\bibfnamefont
  {T.}~\bibnamefont {Kiss}}, \bibinfo {author} {\bibfnamefont {S.}~\bibnamefont
  {Watanabe}}, \bibinfo {author} {\bibfnamefont {C.-T.}\ \bibnamefont {Chen}},
  \bibinfo {author} {\bibfnamefont {K.}~\bibnamefont {Kihou}}, \bibinfo
  {author} {\bibfnamefont {C.~H.}\ \bibnamefont {Lee}}, \bibinfo {author}
  {\bibfnamefont {A.}~\bibnamefont {Iyo}}, \bibinfo {author} {\bibfnamefont
  {H.}~\bibnamefont {Eisaki}}, \bibinfo {author} {\bibfnamefont
  {T.}~\bibnamefont {Saito}}, \bibinfo {author} {\bibfnamefont
  {H.}~\bibnamefont {Fukazawa}}, \bibinfo {author} {\bibfnamefont
  {Y.}~\bibnamefont {Kohori}}, \bibinfo {author} {\bibfnamefont
  {K.}~\bibnamefont {Hashimoto}}, \bibinfo {author} {\bibfnamefont
  {T.}~\bibnamefont {Shibauchi}}, \bibinfo {author} {\bibfnamefont
  {Y.}~\bibnamefont {Matsuda}}, \bibinfo {author} {\bibfnamefont
  {H.}~\bibnamefont {Ikeda}}, \bibinfo {author} {\bibfnamefont
  {H.}~\bibnamefont {Miyahara}}, \bibinfo {author} {\bibfnamefont
  {R.}~\bibnamefont {Arita}}, \bibinfo {author} {\bibfnamefont
  {A.}~\bibnamefont {Chainani}}, \ and\ \bibinfo {author} {\bibfnamefont
  {S.}~\bibnamefont {Shin}},\ }\href {\doibase 10.1126/science.1222793}
  {\bibfield  {journal} {\bibinfo  {journal} {Science}\ }\textbf {\bibinfo
  {volume} {337}},\ \bibinfo {pages} {1314} (\bibinfo {year}
  {2012})}\BibitemShut {NoStop}%
\bibitem [{\citenamefont {Thomale}\ \emph {et~al.}(2011)\citenamefont
  {Thomale}, \citenamefont {Platt}, \citenamefont {Hanke}, \citenamefont {Hu},\
  and\ \citenamefont {Bernevig}}]{Thomale11PRL}%
  \BibitemOpen
  \bibfield  {author} {\bibinfo {author} {\bibfnamefont {R.}~\bibnamefont
  {Thomale}}, \bibinfo {author} {\bibfnamefont {C.}~\bibnamefont {Platt}},
  \bibinfo {author} {\bibfnamefont {W.}~\bibnamefont {Hanke}}, \bibinfo
  {author} {\bibfnamefont {J.}~\bibnamefont {Hu}}, \ and\ \bibinfo {author}
  {\bibfnamefont {B.~A.}\ \bibnamefont {Bernevig}},\ }\href {\doibase
  10.1103/PhysRevLett.107.117001} {\bibfield  {journal} {\bibinfo  {journal}
  {Phys. Rev. Lett.}\ }\textbf {\bibinfo {volume} {107}},\ \bibinfo {pages}
  {117001} (\bibinfo {year} {2011})}\BibitemShut {NoStop}%
\bibitem [{\citenamefont {Maiti}\ \emph {et~al.}(2011)\citenamefont {Maiti},
  \citenamefont {Korshunov}, \citenamefont {Maier}, \citenamefont
  {Hirschfeld},\ and\ \citenamefont {Chubukov}}]{Maiti11PRL}%
  \BibitemOpen
  \bibfield  {author} {\bibinfo {author} {\bibfnamefont {S.}~\bibnamefont
  {Maiti}}, \bibinfo {author} {\bibfnamefont {M.~M.}\ \bibnamefont
  {Korshunov}}, \bibinfo {author} {\bibfnamefont {T.~A.}\ \bibnamefont
  {Maier}}, \bibinfo {author} {\bibfnamefont {P.~J.}\ \bibnamefont
  {Hirschfeld}}, \ and\ \bibinfo {author} {\bibfnamefont {A.~V.}\ \bibnamefont
  {Chubukov}},\ }\href {\doibase 10.1103/PhysRevLett.107.147002} {\bibfield
  {journal} {\bibinfo  {journal} {Phys. Rev. Lett.}\ }\textbf {\bibinfo
  {volume} {107}},\ \bibinfo {pages} {147002} (\bibinfo {year}
  {2011})}\BibitemShut {NoStop}%
\bibitem [{\citenamefont {Suzuki}\ \emph {et~al.}(2011)\citenamefont {Suzuki},
  \citenamefont {Usui},\ and\ \citenamefont {Kuroki}}]{Suzuki11PRB}%
  \BibitemOpen
  \bibfield  {author} {\bibinfo {author} {\bibfnamefont {K.}~\bibnamefont
  {Suzuki}}, \bibinfo {author} {\bibfnamefont {H.}~\bibnamefont {Usui}}, \ and\
  \bibinfo {author} {\bibfnamefont {K.}~\bibnamefont {Kuroki}},\ }\href
  {\doibase 10.1103/PhysRevB.84.144514} {\bibfield  {journal} {\bibinfo
  {journal} {Phys. Rev. B}\ }\textbf {\bibinfo {volume} {84}},\ \bibinfo
  {pages} {144514} (\bibinfo {year} {2011})}\BibitemShut {NoStop}%
\bibitem [{\citenamefont {Terashima}\ \emph
  {et~al.}(2010{\natexlab{b}})\citenamefont {Terashima}, \citenamefont
  {Kimata}, \citenamefont {Kurita}, \citenamefont {Satsukawa}, \citenamefont
  {Harada}, \citenamefont {Hazama}, \citenamefont {Imai}, \citenamefont {Sato},
  \citenamefont {Kihou}, \citenamefont {Lee}, \citenamefont {Kito},
  \citenamefont {Eisaki}, \citenamefont {Iyo}, \citenamefont {Saito},
  \citenamefont {Fukazawa}, \citenamefont {Kohori}, \citenamefont {Harima},\
  and\ \citenamefont {Uji}}]{Terashima10JPSJ}%
  \BibitemOpen
  \bibfield  {author} {\bibinfo {author} {\bibfnamefont {T.}~\bibnamefont
  {Terashima}}, \bibinfo {author} {\bibfnamefont {M.}~\bibnamefont {Kimata}},
  \bibinfo {author} {\bibfnamefont {N.}~\bibnamefont {Kurita}}, \bibinfo
  {author} {\bibfnamefont {H.}~\bibnamefont {Satsukawa}}, \bibinfo {author}
  {\bibfnamefont {A.}~\bibnamefont {Harada}}, \bibinfo {author} {\bibfnamefont
  {K.}~\bibnamefont {Hazama}}, \bibinfo {author} {\bibfnamefont
  {M.}~\bibnamefont {Imai}}, \bibinfo {author} {\bibfnamefont {A.}~\bibnamefont
  {Sato}}, \bibinfo {author} {\bibfnamefont {K.}~\bibnamefont {Kihou}},
  \bibinfo {author} {\bibfnamefont {C.~H.}\ \bibnamefont {Lee}}, \bibinfo
  {author} {\bibfnamefont {H.}~\bibnamefont {Kito}}, \bibinfo {author}
  {\bibfnamefont {H.}~\bibnamefont {Eisaki}}, \bibinfo {author} {\bibfnamefont
  {A.}~\bibnamefont {Iyo}}, \bibinfo {author} {\bibfnamefont {T.}~\bibnamefont
  {Saito}}, \bibinfo {author} {\bibfnamefont {H.}~\bibnamefont {Fukazawa}},
  \bibinfo {author} {\bibfnamefont {Y.}~\bibnamefont {Kohori}}, \bibinfo
  {author} {\bibfnamefont {H.}~\bibnamefont {Harima}}, \ and\ \bibinfo {author}
  {\bibfnamefont {S.}~\bibnamefont {Uji}},\ }\href {\doibase
  10.1143/JPSJ.79.053702} {\bibfield  {journal} {\bibinfo  {journal} {J. Phys.
  Soc. Jpn.}\ }\textbf {\bibinfo {volume} {79}},\ \bibinfo {pages} {053702}
  (\bibinfo {year} {2010}{\natexlab{b}})}\BibitemShut {NoStop}%
\bibitem [{\citenamefont {Sato}\ \emph {et~al.}(2009)\citenamefont {Sato},
  \citenamefont {Nakayama}, \citenamefont {Sekiba}, \citenamefont {Richard},
  \citenamefont {Xu}, \citenamefont {Souma}, \citenamefont {Takahashi},
  \citenamefont {Chen}, \citenamefont {Luo}, \citenamefont {Wang},\ and\
  \citenamefont {Ding}}]{Sato09PRL}%
  \BibitemOpen
  \bibfield  {author} {\bibinfo {author} {\bibfnamefont {T.}~\bibnamefont
  {Sato}}, \bibinfo {author} {\bibfnamefont {K.}~\bibnamefont {Nakayama}},
  \bibinfo {author} {\bibfnamefont {Y.}~\bibnamefont {Sekiba}}, \bibinfo
  {author} {\bibfnamefont {P.}~\bibnamefont {Richard}}, \bibinfo {author}
  {\bibfnamefont {Y.-M.}\ \bibnamefont {Xu}}, \bibinfo {author} {\bibfnamefont
  {S.}~\bibnamefont {Souma}}, \bibinfo {author} {\bibfnamefont
  {T.}~\bibnamefont {Takahashi}}, \bibinfo {author} {\bibfnamefont {G.~F.}\
  \bibnamefont {Chen}}, \bibinfo {author} {\bibfnamefont {J.~L.}\ \bibnamefont
  {Luo}}, \bibinfo {author} {\bibfnamefont {N.~L.}\ \bibnamefont {Wang}}, \
  and\ \bibinfo {author} {\bibfnamefont {H.}~\bibnamefont {Ding}},\ }\href
  {\doibase 10.1103/PhysRevLett.103.047002} {\bibfield  {journal} {\bibinfo
  {journal} {Phys. Rev. Lett.}\ }\textbf {\bibinfo {volume} {103}},\ \bibinfo
  {eid} {047002} (\bibinfo {year} {2009})}\BibitemShut {NoStop}%
\bibitem [{\citenamefont {Shoenberg}(1984)}]{Shoenberg84}%
  \BibitemOpen
  \bibfield  {author} {\bibinfo {author} {\bibfnamefont {D.}~\bibnamefont
  {Shoenberg}},\ }\href@noop {} {\emph {\bibinfo {title} {Magnetic oscillations
  in metals}}}\ (\bibinfo  {publisher} {Cambridge University Press},\ \bibinfo
  {address} {Cambridge},\ \bibinfo {year} {1984})\BibitemShut {NoStop}%
\bibitem [{Note1()}]{Note1}%
  \BibitemOpen
  \bibinfo {note} {For a 2D system, the number of states is given by $N=(V/4\pi
  ^3)A_0t$, where $A_0$ is the FS cross section normal to the $c$ axis and $t$
  is the thickness of the Brillouin zone along the $c$ axis. Since
  $m^*_0=({\mathchar '26\mkern -9muh}^2/2\pi )\partial A_0/\partial E$, the
  density of states $D=\partial N/\partial E$ is proportional to $m_0^*$ but
  does not depend on the FS size.}\BibitemShut {Stop}%
\bibitem [{\citenamefont {Yamaji}(1989)}]{Yamaji89JPSJ}%
  \BibitemOpen
  \bibfield  {author} {\bibinfo {author} {\bibfnamefont {K.}~\bibnamefont
  {Yamaji}},\ }\href {\doibase 10.1143/JPSJ.58.1520} {\bibfield  {journal}
  {\bibinfo  {journal} {J. Phys. Soc. Jpn.}\ }\textbf {\bibinfo {volume}
  {58}},\ \bibinfo {pages} {1520} (\bibinfo {year} {1989})}\BibitemShut
  {NoStop}%
\bibitem [{Note2()}]{Note2}%
  \BibitemOpen
  \bibinfo {note} {Pick-up coil signals detected at the modulation frequency
  are proportional to ac magnetic susceptibilities for $B \parallel c$. In the
  case of $B \parallel ab$, the coil axis is perpendicular to $B$, since the
  coil and sample are rotated together with the $c$ axis kept parallel to the
  coil axis. If the alignment of the sample and coil axis is perfect, no signal
  will be detected. However, there is small misalignment, which produces small
  signal proportional to the component of ac magnetization along the coil axis
  for $B \parallel ab$.}\BibitemShut {Stop}%
\bibitem [{\citenamefont {Terashima}\ \emph {et~al.}(2009)\citenamefont
  {Terashima}, \citenamefont {Kimata}, \citenamefont {Satsukawa}, \citenamefont
  {Harada}, \citenamefont {Hazama}, \citenamefont {Uji}, \citenamefont
  {Harima}, \citenamefont {Chen}, \citenamefont {Luo},\ and\ \citenamefont
  {Wang}}]{Terashima09JPSJKFA}%
  \BibitemOpen
  \bibfield  {author} {\bibinfo {author} {\bibfnamefont {T.}~\bibnamefont
  {Terashima}}, \bibinfo {author} {\bibfnamefont {M.}~\bibnamefont {Kimata}},
  \bibinfo {author} {\bibfnamefont {H.}~\bibnamefont {Satsukawa}}, \bibinfo
  {author} {\bibfnamefont {A.}~\bibnamefont {Harada}}, \bibinfo {author}
  {\bibfnamefont {K.}~\bibnamefont {Hazama}}, \bibinfo {author} {\bibfnamefont
  {S.}~\bibnamefont {Uji}}, \bibinfo {author} {\bibfnamefont {H.}~\bibnamefont
  {Harima}}, \bibinfo {author} {\bibfnamefont {G.-F.}\ \bibnamefont {Chen}},
  \bibinfo {author} {\bibfnamefont {J.-L.}\ \bibnamefont {Luo}}, \ and\
  \bibinfo {author} {\bibfnamefont {N.-L.}\ \bibnamefont {Wang}},\ }\href
  {\doibase 10.1143/JPSJ.78.063702} {\bibfield  {journal} {\bibinfo  {journal}
  {J. Phys. Soc. Jpn.}\ }\textbf {\bibinfo {volume} {78}},\ \bibinfo {pages}
  {063702} (\bibinfo {year} {2009})}\BibitemShut {NoStop}%
\bibitem [{\citenamefont {Burger}\ \emph {et~al.}(2013)\citenamefont {Burger},
  \citenamefont {Hardy}, \citenamefont {Aoki}, \citenamefont {B\"ohmer},
  \citenamefont {Heid}, \citenamefont {Wolf}, \citenamefont {Schweiss},
  \citenamefont {Fromknecht}, \citenamefont {Jackson}, \citenamefont
  {Paulsen},\ and\ \citenamefont {Meingast}}]{Burger13condmat}%
  \BibitemOpen
  \bibfield  {author} {\bibinfo {author} {\bibfnamefont {P.}~\bibnamefont
  {Burger}}, \bibinfo {author} {\bibfnamefont {F.}~\bibnamefont {Hardy}},
  \bibinfo {author} {\bibfnamefont {D.}~\bibnamefont {Aoki}}, \bibinfo {author}
  {\bibfnamefont {A.~E.}\ \bibnamefont {B\"ohmer}}, \bibinfo {author}
  {\bibfnamefont {R.}~\bibnamefont {Heid}}, \bibinfo {author} {\bibfnamefont
  {T.}~\bibnamefont {Wolf}}, \bibinfo {author} {\bibfnamefont {P.}~\bibnamefont
  {Schweiss}}, \bibinfo {author} {\bibfnamefont {R.}~\bibnamefont
  {Fromknecht}}, \bibinfo {author} {\bibfnamefont {M.~J.}\ \bibnamefont
  {Jackson}}, \bibinfo {author} {\bibfnamefont {C.}~\bibnamefont {Paulsen}}, \
  and\ \bibinfo {author} {\bibfnamefont {C.}~\bibnamefont {Meingast}},\
  }\href@noop {} {\bibfield  {journal} {\bibinfo  {journal} {arXiv:1303.6822}\
  } (\bibinfo {year} {2013})}\BibitemShut {NoStop}%
\bibitem [{\citenamefont {Kim}\ \emph {et~al.}(2011)\citenamefont {Kim},
  \citenamefont {Kim}, \citenamefont {Stewart}, \citenamefont {Chen},\ and\
  \citenamefont {Wang}}]{Kim_C11PRB}%
  \BibitemOpen
  \bibfield  {author} {\bibinfo {author} {\bibfnamefont {J.~S.}\ \bibnamefont
  {Kim}}, \bibinfo {author} {\bibfnamefont {E.~G.}\ \bibnamefont {Kim}},
  \bibinfo {author} {\bibfnamefont {G.~R.}\ \bibnamefont {Stewart}}, \bibinfo
  {author} {\bibfnamefont {X.~H.}\ \bibnamefont {Chen}}, \ and\ \bibinfo
  {author} {\bibfnamefont {X.~F.}\ \bibnamefont {Wang}},\ }\href {\doibase
  10.1103/PhysRevB.83.172502} {\bibfield  {journal} {\bibinfo  {journal} {Phys.
  Rev. B}\ }\textbf {\bibinfo {volume} {83}},\ \bibinfo {pages} {172502}
  (\bibinfo {year} {2011})}\BibitemShut {NoStop}%
\bibitem [{\citenamefont {Abdel-Hafiez}\ \emph {et~al.}(2012)\citenamefont
  {Abdel-Hafiez}, \citenamefont {Aswartham}, \citenamefont {Wurmehl},
  \citenamefont {Grinenko}, \citenamefont {Hess}, \citenamefont {Drechsler},
  \citenamefont {Johnston}, \citenamefont {Wolter}, \citenamefont {B\"uchner},
  \citenamefont {Rosner},\ and\ \citenamefont {Boeri}}]{Abdel-Hafiez12PRB}%
  \BibitemOpen
  \bibfield  {author} {\bibinfo {author} {\bibfnamefont {M.}~\bibnamefont
  {Abdel-Hafiez}}, \bibinfo {author} {\bibfnamefont {S.}~\bibnamefont
  {Aswartham}}, \bibinfo {author} {\bibfnamefont {S.}~\bibnamefont {Wurmehl}},
  \bibinfo {author} {\bibfnamefont {V.}~\bibnamefont {Grinenko}}, \bibinfo
  {author} {\bibfnamefont {C.}~\bibnamefont {Hess}}, \bibinfo {author}
  {\bibfnamefont {S.-L.}\ \bibnamefont {Drechsler}}, \bibinfo {author}
  {\bibfnamefont {S.}~\bibnamefont {Johnston}}, \bibinfo {author}
  {\bibfnamefont {A.~U.~B.}\ \bibnamefont {Wolter}}, \bibinfo {author}
  {\bibfnamefont {B.}~\bibnamefont {B\"uchner}}, \bibinfo {author}
  {\bibfnamefont {H.}~\bibnamefont {Rosner}}, \ and\ \bibinfo {author}
  {\bibfnamefont {L.}~\bibnamefont {Boeri}},\ }\href {\doibase
  10.1103/PhysRevB.85.134533} {\bibfield  {journal} {\bibinfo  {journal} {Phys.
  Rev. B}\ }\textbf {\bibinfo {volume} {85}},\ \bibinfo {pages} {134533}
  (\bibinfo {year} {2012})}\BibitemShut {NoStop}%
\bibitem [{\citenamefont {Terashima}\ \emph {et~al.}(2013)\citenamefont
  {Terashima}, \citenamefont {Kihou}, \citenamefont {Tomita}, \citenamefont
  {Tsuchiya}, \citenamefont {Kikugawa}, \citenamefont {Ishida}, \citenamefont
  {Lee}, \citenamefont {Iyo}, \citenamefont {Eisaki},\ and\ \citenamefont
  {Uji}}]{Terashima13PRB}%
  \BibitemOpen
  \bibfield  {author} {\bibinfo {author} {\bibfnamefont {T.}~\bibnamefont
  {Terashima}}, \bibinfo {author} {\bibfnamefont {K.}~\bibnamefont {Kihou}},
  \bibinfo {author} {\bibfnamefont {M.}~\bibnamefont {Tomita}}, \bibinfo
  {author} {\bibfnamefont {S.}~\bibnamefont {Tsuchiya}}, \bibinfo {author}
  {\bibfnamefont {N.}~\bibnamefont {Kikugawa}}, \bibinfo {author}
  {\bibfnamefont {S.}~\bibnamefont {Ishida}}, \bibinfo {author} {\bibfnamefont
  {C.~H.}\ \bibnamefont {Lee}}, \bibinfo {author} {\bibfnamefont
  {A.}~\bibnamefont {Iyo}}, \bibinfo {author} {\bibfnamefont {H.}~\bibnamefont
  {Eisaki}}, \ and\ \bibinfo {author} {\bibfnamefont {S.}~\bibnamefont {Uji}},\
  }\href {\doibase 10.1103/PhysRevB.87.184513} {\bibfield  {journal} {\bibinfo
  {journal} {Phys. Rev. B}\ }\textbf {\bibinfo {volume} {87}},\ \bibinfo
  {pages} {184513} (\bibinfo {year} {2013})}\BibitemShut {NoStop}%
\bibitem [{Note3()}]{Note3}%
  \BibitemOpen
  \bibinfo {note} {This is related to the imbalance of the pick-up coil. The
  pick-up coil is composed of coaxially wound inner and outer coils. The two
  coils are balanced without a sample at $\theta $ = 0 so that the emf's
  induced in the two coils cancel each other out. The balance however degrades
  as $\theta $ increases.}\BibitemShut {Stop}%
\bibitem [{\citenamefont {Yoshida}\ \emph {et~al.}(2011)\citenamefont
  {Yoshida}, \citenamefont {Nishi}, \citenamefont {Fujimori}, \citenamefont
  {Yi}, \citenamefont {Moore}, \citenamefont {Lu}, \citenamefont {Shen},
  \citenamefont {Kihou}, \citenamefont {Shirage}, \citenamefont {Kito},
  \citenamefont {Lee}, \citenamefont {Iyo}, \citenamefont {Eisaki},\ and\
  \citenamefont {Harima}}]{Yoshida11JPCS}%
  \BibitemOpen
  \bibfield  {author} {\bibinfo {author} {\bibfnamefont {T.}~\bibnamefont
  {Yoshida}}, \bibinfo {author} {\bibfnamefont {I.}~\bibnamefont {Nishi}},
  \bibinfo {author} {\bibfnamefont {A.}~\bibnamefont {Fujimori}}, \bibinfo
  {author} {\bibfnamefont {M.}~\bibnamefont {Yi}}, \bibinfo {author}
  {\bibfnamefont {R.~G.}\ \bibnamefont {Moore}}, \bibinfo {author}
  {\bibfnamefont {D.-H.}\ \bibnamefont {Lu}}, \bibinfo {author} {\bibfnamefont
  {Z.-X.}\ \bibnamefont {Shen}}, \bibinfo {author} {\bibfnamefont
  {K.}~\bibnamefont {Kihou}}, \bibinfo {author} {\bibfnamefont {P.~M.}\
  \bibnamefont {Shirage}}, \bibinfo {author} {\bibfnamefont {H.}~\bibnamefont
  {Kito}}, \bibinfo {author} {\bibfnamefont {C.~H.}\ \bibnamefont {Lee}},
  \bibinfo {author} {\bibfnamefont {A.}~\bibnamefont {Iyo}}, \bibinfo {author}
  {\bibfnamefont {H.}~\bibnamefont {Eisaki}}, \ and\ \bibinfo {author}
  {\bibfnamefont {H.}~\bibnamefont {Harima}},\ }\href {\doibase
  10.1016/j.jpcs.2010.10.064} {\bibfield  {journal} {\bibinfo  {journal} {J.
  Phys. Chem. Solids}\ }\textbf {\bibinfo {volume} {72}},\ \bibinfo {pages}
  {465 } (\bibinfo {year} {2011})}\BibitemShut {NoStop}%
\bibitem [{\citenamefont {Wray}\ \emph {et~al.}(2012)\citenamefont {Wray},
  \citenamefont {Thomale}, \citenamefont {Platt}, \citenamefont {Hsieh},
  \citenamefont {Qian}, \citenamefont {Chen}, \citenamefont {Luo},
  \citenamefont {Wang},\ and\ \citenamefont {Hasan}}]{Wray12PRB}%
  \BibitemOpen
  \bibfield  {author} {\bibinfo {author} {\bibfnamefont {L.~A.}\ \bibnamefont
  {Wray}}, \bibinfo {author} {\bibfnamefont {R.}~\bibnamefont {Thomale}},
  \bibinfo {author} {\bibfnamefont {C.}~\bibnamefont {Platt}}, \bibinfo
  {author} {\bibfnamefont {D.}~\bibnamefont {Hsieh}}, \bibinfo {author}
  {\bibfnamefont {D.}~\bibnamefont {Qian}}, \bibinfo {author} {\bibfnamefont
  {G.~F.}\ \bibnamefont {Chen}}, \bibinfo {author} {\bibfnamefont {J.~L.}\
  \bibnamefont {Luo}}, \bibinfo {author} {\bibfnamefont {N.~L.}\ \bibnamefont
  {Wang}}, \ and\ \bibinfo {author} {\bibfnamefont {M.~Z.}\ \bibnamefont
  {Hasan}},\ }\href {\doibase 10.1103/PhysRevB.86.144515} {\bibfield  {journal}
  {\bibinfo  {journal} {Phys. Rev. B}\ }\textbf {\bibinfo {volume} {86}},\
  \bibinfo {pages} {144515} (\bibinfo {year} {2012})}\BibitemShut {NoStop}%
\bibitem [{\citenamefont {Yoshida}\ \emph {et~al.}(2012)\citenamefont
  {Yoshida}, \citenamefont {Ideta}, \citenamefont {Nishi}, \citenamefont
  {Fujimori}, \citenamefont {Yi}, \citenamefont {Moore}, \citenamefont {Mo},
  \citenamefont {Lu}, \citenamefont {Shen}, \citenamefont {Hussain},
  \citenamefont {Kihou}, \citenamefont {Shirage}, \citenamefont {Kito},
  \citenamefont {Lee}, \citenamefont {Iyo}, \citenamefont {Eisaki},\ and\
  \citenamefont {Harima}}]{Yoshida12condmat}%
  \BibitemOpen
  \bibfield  {author} {\bibinfo {author} {\bibfnamefont {T.}~\bibnamefont
  {Yoshida}}, \bibinfo {author} {\bibfnamefont {S.}~\bibnamefont {Ideta}},
  \bibinfo {author} {\bibfnamefont {I.}~\bibnamefont {Nishi}}, \bibinfo
  {author} {\bibfnamefont {A.}~\bibnamefont {Fujimori}}, \bibinfo {author}
  {\bibfnamefont {M.}~\bibnamefont {Yi}}, \bibinfo {author} {\bibfnamefont
  {R.~G.}\ \bibnamefont {Moore}}, \bibinfo {author} {\bibfnamefont {S.~K.}\
  \bibnamefont {Mo}}, \bibinfo {author} {\bibfnamefont {D.-H.}\ \bibnamefont
  {Lu}}, \bibinfo {author} {\bibfnamefont {Z.-X.}\ \bibnamefont {Shen}},
  \bibinfo {author} {\bibfnamefont {Z.}~\bibnamefont {Hussain}}, \bibinfo
  {author} {\bibfnamefont {K.}~\bibnamefont {Kihou}}, \bibinfo {author}
  {\bibfnamefont {P.~M.}\ \bibnamefont {Shirage}}, \bibinfo {author}
  {\bibfnamefont {H.}~\bibnamefont {Kito}}, \bibinfo {author} {\bibfnamefont
  {C.~H.}\ \bibnamefont {Lee}}, \bibinfo {author} {\bibfnamefont
  {A.}~\bibnamefont {Iyo}}, \bibinfo {author} {\bibfnamefont {H.}~\bibnamefont
  {Eisaki}}, \ and\ \bibinfo {author} {\bibfnamefont {H.}~\bibnamefont
  {Harima}},\ }\href@noop {} {\bibfield  {journal} {\bibinfo  {journal}
  {arXiv:1205.6911}\ } (\bibinfo {year} {2012})}\BibitemShut {NoStop}%
\bibitem [{\citenamefont {Kimata}\ \emph {et~al.}(2010)\citenamefont {Kimata},
  \citenamefont {Terashima}, \citenamefont {Kurita}, \citenamefont {Satsukawa},
  \citenamefont {Harada}, \citenamefont {Kodama}, \citenamefont {Sato},
  \citenamefont {Imai}, \citenamefont {Kihou}, \citenamefont {Lee},
  \citenamefont {Kito}, \citenamefont {Eisaki}, \citenamefont {Iyo},
  \citenamefont {Saito}, \citenamefont {Fukazawa}, \citenamefont {Kohori},
  \citenamefont {Harima},\ and\ \citenamefont {Uji}}]{Kimata10PRL}%
  \BibitemOpen
  \bibfield  {author} {\bibinfo {author} {\bibfnamefont {M.}~\bibnamefont
  {Kimata}}, \bibinfo {author} {\bibfnamefont {T.}~\bibnamefont {Terashima}},
  \bibinfo {author} {\bibfnamefont {N.}~\bibnamefont {Kurita}}, \bibinfo
  {author} {\bibfnamefont {H.}~\bibnamefont {Satsukawa}}, \bibinfo {author}
  {\bibfnamefont {A.}~\bibnamefont {Harada}}, \bibinfo {author} {\bibfnamefont
  {K.}~\bibnamefont {Kodama}}, \bibinfo {author} {\bibfnamefont
  {A.}~\bibnamefont {Sato}}, \bibinfo {author} {\bibfnamefont {M.}~\bibnamefont
  {Imai}}, \bibinfo {author} {\bibfnamefont {K.}~\bibnamefont {Kihou}},
  \bibinfo {author} {\bibfnamefont {C.~H.}\ \bibnamefont {Lee}}, \bibinfo
  {author} {\bibfnamefont {H.}~\bibnamefont {Kito}}, \bibinfo {author}
  {\bibfnamefont {H.}~\bibnamefont {Eisaki}}, \bibinfo {author} {\bibfnamefont
  {A.}~\bibnamefont {Iyo}}, \bibinfo {author} {\bibfnamefont {T.}~\bibnamefont
  {Saito}}, \bibinfo {author} {\bibfnamefont {H.}~\bibnamefont {Fukazawa}},
  \bibinfo {author} {\bibfnamefont {Y.}~\bibnamefont {Kohori}}, \bibinfo
  {author} {\bibfnamefont {H.}~\bibnamefont {Harima}}, \ and\ \bibinfo {author}
  {\bibfnamefont {S.}~\bibnamefont {Uji}},\ }\href {\doibase
  10.1103/PhysRevLett.105.246403} {\bibfield  {journal} {\bibinfo  {journal}
  {Phys. Rev. Lett.}\ }\textbf {\bibinfo {volume} {105}},\ \bibinfo {pages}
  {246403} (\bibinfo {year} {2010})}\BibitemShut {NoStop}%
\bibitem [{Note4()}]{Note4}%
  \BibitemOpen
  \bibinfo {note} {Because $m^*$ varies as $\sim 1/\mathop {\mathgroup
  \symoperators cos}\nolimits \theta $, $S$ in Eq. (5) varies with $\theta
  $.}\BibitemShut {Stop}%
\bibitem [{\citenamefont {Yanase}(1985)}]{Yanase1995}%
  \BibitemOpen
  \bibfield  {author} {\bibinfo {author} {\bibfnamefont {A.}~\bibnamefont
  {Yanase}},\ }\href@noop {} {\emph {\bibinfo {title} {Fortran Program for
  Space Group}}},\ \bibinfo {edition} {1st}\ ed.\ (\bibinfo  {publisher}
  {Shokabo},\ \bibinfo {address} {Tokyo},\ \bibinfo {year} {1985})\ \bibinfo
  {note} {[in Japanese]}\BibitemShut {NoStop}%
\bibitem [{\citenamefont {Rozsa}\ and\ \citenamefont
  {Schuster}(1981)}]{Rozsa81ZNB}%
  \BibitemOpen
  \bibfield  {author} {\bibinfo {author} {\bibfnamefont {S.}~\bibnamefont
  {Rozsa}}\ and\ \bibinfo {author} {\bibfnamefont {H.~U.}\ \bibnamefont
  {Schuster}},\ }\href@noop {} {\bibfield  {journal} {\bibinfo  {journal} {Z.
  Naturforsch. B}\ }\textbf {\bibinfo {volume} {36}},\ \bibinfo {pages} {1668}
  (\bibinfo {year} {1981})}\BibitemShut {NoStop}%
\bibitem [{\citenamefont {Fukazawa}\ \emph {et~al.}(2011)\citenamefont
  {Fukazawa}, \citenamefont {Saito}, \citenamefont {Yamada}, \citenamefont
  {Kondo}, \citenamefont {Hirano}, \citenamefont {Kohori}, \citenamefont
  {Kuga}, \citenamefont {Sakai}, \citenamefont {Matsumoto}, \citenamefont
  {Nakatsuji}, \citenamefont {Kihou}, \citenamefont {Iyo}, \citenamefont
  {Lee},\ and\ \citenamefont {Eisaki}}]{Fukazawa11JPSJ_SA}%
  \BibitemOpen
  \bibfield  {author} {\bibinfo {author} {\bibfnamefont {H.}~\bibnamefont
  {Fukazawa}}, \bibinfo {author} {\bibfnamefont {T.}~\bibnamefont {Saito}},
  \bibinfo {author} {\bibfnamefont {Y.}~\bibnamefont {Yamada}}, \bibinfo
  {author} {\bibfnamefont {K.}~\bibnamefont {Kondo}}, \bibinfo {author}
  {\bibfnamefont {M.}~\bibnamefont {Hirano}}, \bibinfo {author} {\bibfnamefont
  {Y.}~\bibnamefont {Kohori}}, \bibinfo {author} {\bibfnamefont
  {K.}~\bibnamefont {Kuga}}, \bibinfo {author} {\bibfnamefont {A.}~\bibnamefont
  {Sakai}}, \bibinfo {author} {\bibfnamefont {Y.}~\bibnamefont {Matsumoto}},
  \bibinfo {author} {\bibfnamefont {S.}~\bibnamefont {Nakatsuji}}, \bibinfo
  {author} {\bibfnamefont {K.}~\bibnamefont {Kihou}}, \bibinfo {author}
  {\bibfnamefont {A.}~\bibnamefont {Iyo}}, \bibinfo {author} {\bibfnamefont
  {C.~H.}\ \bibnamefont {Lee}}, \ and\ \bibinfo {author} {\bibfnamefont
  {H.}~\bibnamefont {Eisaki}},\ }\href {\doibase 10.1143/JPSJS.80SA.SA118}
  {\bibfield  {journal} {\bibinfo  {journal} {J. Phys. Soc. Jpn.}\ }\textbf
  {\bibinfo {volume} {80SA}},\ \bibinfo {pages} {SA118} (\bibinfo {year}
  {2011})}\BibitemShut {NoStop}%
\bibitem [{\citenamefont {Singh}(2009)}]{Singh09PRB}%
  \BibitemOpen
  \bibfield  {author} {\bibinfo {author} {\bibfnamefont {D.~J.}\ \bibnamefont
  {Singh}},\ }\href {\doibase 10.1103/PhysRevB.79.174520} {\bibfield  {journal}
  {\bibinfo  {journal} {Phys. Rev. B}\ }\textbf {\bibinfo {volume} {79}},\
  \bibinfo {pages} {174520} (\bibinfo {year} {2009})}\BibitemShut {NoStop}%
\bibitem [{Note5()}]{Note5}%
  \BibitemOpen
  \bibinfo {note} {The adjusted calculation gives a Sommerfeld coefficient of
  33.1 mJ/K$^2$mol. However, the validity of this estimate is questioned
  because the adjustment can not reproduce the observed FS.}\BibitemShut
  {Stop}%
\bibitem [{\citenamefont {Haule}\ \emph {et~al.}(2008)\citenamefont {Haule},
  \citenamefont {Shim},\ and\ \citenamefont {Kotliar}}]{Haule08PRL}%
  \BibitemOpen
  \bibfield  {author} {\bibinfo {author} {\bibfnamefont {K.}~\bibnamefont
  {Haule}}, \bibinfo {author} {\bibfnamefont {J.~H.}\ \bibnamefont {Shim}}, \
  and\ \bibinfo {author} {\bibfnamefont {G.}~\bibnamefont {Kotliar}},\ }\href
  {\doibase 10.1103/PhysRevLett.100.226402} {\bibfield  {journal} {\bibinfo
  {journal} {Phys. Rev. Lett.}\ }\textbf {\bibinfo {volume} {100}},\ \bibinfo
  {pages} {226402} (\bibinfo {year} {2008})}\BibitemShut {NoStop}%
\bibitem [{\citenamefont {Aichhorn}\ \emph {et~al.}(2009)\citenamefont
  {Aichhorn}, \citenamefont {Pourovskii}, \citenamefont {Vildosola},
  \citenamefont {Ferrero}, \citenamefont {Parcollet}, \citenamefont {Miyake},
  \citenamefont {Georges},\ and\ \citenamefont {Biermann}}]{Aichhorn09PRB}%
  \BibitemOpen
  \bibfield  {author} {\bibinfo {author} {\bibfnamefont {M.}~\bibnamefont
  {Aichhorn}}, \bibinfo {author} {\bibfnamefont {L.}~\bibnamefont
  {Pourovskii}}, \bibinfo {author} {\bibfnamefont {V.}~\bibnamefont
  {Vildosola}}, \bibinfo {author} {\bibfnamefont {M.}~\bibnamefont {Ferrero}},
  \bibinfo {author} {\bibfnamefont {O.}~\bibnamefont {Parcollet}}, \bibinfo
  {author} {\bibfnamefont {T.}~\bibnamefont {Miyake}}, \bibinfo {author}
  {\bibfnamefont {A.}~\bibnamefont {Georges}}, \ and\ \bibinfo {author}
  {\bibfnamefont {S.}~\bibnamefont {Biermann}},\ }\href {\doibase
  10.1103/PhysRevB.80.085101} {\bibfield  {journal} {\bibinfo  {journal} {Phys.
  Rev. B}\ }\textbf {\bibinfo {volume} {80}},\ \bibinfo {eid} {085101}
  (\bibinfo {year} {2009})}\BibitemShut {NoStop}%
\bibitem [{\citenamefont {Yin}\ \emph {et~al.}(2011)\citenamefont {Yin},
  \citenamefont {Haule},\ and\ \citenamefont {Kotliar}}]{Yin11NatMat}%
  \BibitemOpen
  \bibfield  {author} {\bibinfo {author} {\bibfnamefont {Z.~P.}\ \bibnamefont
  {Yin}}, \bibinfo {author} {\bibfnamefont {K.}~\bibnamefont {Haule}}, \ and\
  \bibinfo {author} {\bibfnamefont {G.}~\bibnamefont {Kotliar}},\ }\href
  {http://dx.doi.org/10.1038/nmat3120} {\bibfield  {journal} {\bibinfo
  {journal} {Nat Mater}\ }\textbf {\bibinfo {volume} {10}},\ \bibinfo {pages}
  {932} (\bibinfo {year} {2011})}\BibitemShut {NoStop}%
\bibitem [{\citenamefont {Hardy}\ \emph {et~al.}(2013)\citenamefont {Hardy},
  \citenamefont {Boehmer}, \citenamefont {Aoki}, \citenamefont {Burger},
  \citenamefont {Wolf}, \citenamefont {Schweiss}, \citenamefont {Heid},
  \citenamefont {Adelmann}, \citenamefont {Yao}, \citenamefont {Kotliar},
  \citenamefont {Schmalian},\ and\ \citenamefont {Meingast}}]{Hardy13condmat}%
  \BibitemOpen
  \bibfield  {author} {\bibinfo {author} {\bibfnamefont {F.}~\bibnamefont
  {Hardy}}, \bibinfo {author} {\bibfnamefont {A.~E.}\ \bibnamefont {Boehmer}},
  \bibinfo {author} {\bibfnamefont {D.}~\bibnamefont {Aoki}}, \bibinfo {author}
  {\bibfnamefont {P.}~\bibnamefont {Burger}}, \bibinfo {author} {\bibfnamefont
  {T.}~\bibnamefont {Wolf}}, \bibinfo {author} {\bibfnamefont {P.}~\bibnamefont
  {Schweiss}}, \bibinfo {author} {\bibfnamefont {R.}~\bibnamefont {Heid}},
  \bibinfo {author} {\bibfnamefont {P.}~\bibnamefont {Adelmann}}, \bibinfo
  {author} {\bibfnamefont {Y.~X.}\ \bibnamefont {Yao}}, \bibinfo {author}
  {\bibfnamefont {G.}~\bibnamefont {Kotliar}}, \bibinfo {author} {\bibfnamefont
  {J.}~\bibnamefont {Schmalian}}, \ and\ \bibinfo {author} {\bibfnamefont
  {C.}~\bibnamefont {Meingast}},\ }\href@noop {} {\bibfield  {journal}
  {\bibinfo  {journal} {arXiv:1302.1696}\ } (\bibinfo {year}
  {2013})}\BibitemShut {NoStop}%
\bibitem [{\citenamefont {Yi}\ \emph {et~al.}(2009)\citenamefont {Yi},
  \citenamefont {Lu}, \citenamefont {Analytis}, \citenamefont {Chu},
  \citenamefont {Mo}, \citenamefont {He}, \citenamefont {Moore}, \citenamefont
  {Zhou}, \citenamefont {Chen}, \citenamefont {Luo}, \citenamefont {Wang},
  \citenamefont {Hussain}, \citenamefont {Singh}, \citenamefont {Fisher},\ and\
  \citenamefont {Shen}}]{Yi09PRB}%
  \BibitemOpen
  \bibfield  {author} {\bibinfo {author} {\bibfnamefont {M.}~\bibnamefont
  {Yi}}, \bibinfo {author} {\bibfnamefont {D.~H.}\ \bibnamefont {Lu}}, \bibinfo
  {author} {\bibfnamefont {J.~G.}\ \bibnamefont {Analytis}}, \bibinfo {author}
  {\bibfnamefont {J.-H.}\ \bibnamefont {Chu}}, \bibinfo {author} {\bibfnamefont
  {S.-K.}\ \bibnamefont {Mo}}, \bibinfo {author} {\bibfnamefont {R.-H.}\
  \bibnamefont {He}}, \bibinfo {author} {\bibfnamefont {R.~G.}\ \bibnamefont
  {Moore}}, \bibinfo {author} {\bibfnamefont {X.~J.}\ \bibnamefont {Zhou}},
  \bibinfo {author} {\bibfnamefont {G.~F.}\ \bibnamefont {Chen}}, \bibinfo
  {author} {\bibfnamefont {J.~L.}\ \bibnamefont {Luo}}, \bibinfo {author}
  {\bibfnamefont {N.~L.}\ \bibnamefont {Wang}}, \bibinfo {author}
  {\bibfnamefont {Z.}~\bibnamefont {Hussain}}, \bibinfo {author} {\bibfnamefont
  {D.~J.}\ \bibnamefont {Singh}}, \bibinfo {author} {\bibfnamefont {I.~R.}\
  \bibnamefont {Fisher}}, \ and\ \bibinfo {author} {\bibfnamefont {Z.-X.}\
  \bibnamefont {Shen}},\ }\href {\doibase 10.1103/PhysRevB.80.024515}
  {\bibfield  {journal} {\bibinfo  {journal} {Phys. Rev. B}\ }\textbf {\bibinfo
  {volume} {80}},\ \bibinfo {pages} {024515} (\bibinfo {year}
  {2009})}\BibitemShut {NoStop}%
\bibitem [{\citenamefont {Ikeda}\ \emph {et~al.}(2010)\citenamefont {Ikeda},
  \citenamefont {Arita},\ and\ \citenamefont {Kune\ifmmode~\check{s}\else
  \v{s}\fi{}}}]{Ikeda10PRB}%
  \BibitemOpen
  \bibfield  {author} {\bibinfo {author} {\bibfnamefont {H.}~\bibnamefont
  {Ikeda}}, \bibinfo {author} {\bibfnamefont {R.}~\bibnamefont {Arita}}, \ and\
  \bibinfo {author} {\bibfnamefont {J.}~\bibnamefont
  {Kune\ifmmode~\check{s}\else \v{s}\fi{}}},\ }\href {\doibase
  10.1103/PhysRevB.82.024508} {\bibfield  {journal} {\bibinfo  {journal} {Phys.
  Rev. B}\ }\textbf {\bibinfo {volume} {82}},\ \bibinfo {pages} {024508}
  (\bibinfo {year} {2010})}\BibitemShut {NoStop}%
\bibitem [{\citenamefont {Ortenzi}\ \emph {et~al.}(2009)\citenamefont
  {Ortenzi}, \citenamefont {Cappelluti}, \citenamefont {Benfatto},\ and\
  \citenamefont {Pietronero}}]{Ortenzi09PRL}%
  \BibitemOpen
  \bibfield  {author} {\bibinfo {author} {\bibfnamefont {L.}~\bibnamefont
  {Ortenzi}}, \bibinfo {author} {\bibfnamefont {E.}~\bibnamefont {Cappelluti}},
  \bibinfo {author} {\bibfnamefont {L.}~\bibnamefont {Benfatto}}, \ and\
  \bibinfo {author} {\bibfnamefont {L.}~\bibnamefont {Pietronero}},\ }\href
  {\doibase 10.1103/PhysRevLett.103.046404} {\bibfield  {journal} {\bibinfo
  {journal} {Phys. Rev. Lett.}\ }\textbf {\bibinfo {volume} {103}},\ \bibinfo
  {eid} {046404} (\bibinfo {year} {2009})}\BibitemShut {NoStop}%
\bibitem [{\citenamefont {Lee}\ \emph {et~al.}(2011)\citenamefont {Lee},
  \citenamefont {Kihou}, \citenamefont {Kawano-Furukawa}, \citenamefont
  {Saito}, \citenamefont {Iyo}, \citenamefont {Eisaki}, \citenamefont
  {Fukazawa}, \citenamefont {Kohori}, \citenamefont {Suzuki}, \citenamefont
  {Usui}, \citenamefont {Kuroki},\ and\ \citenamefont {Yamada}}]{Lee11PRL}%
  \BibitemOpen
  \bibfield  {author} {\bibinfo {author} {\bibfnamefont {C.~H.}\ \bibnamefont
  {Lee}}, \bibinfo {author} {\bibfnamefont {K.}~\bibnamefont {Kihou}}, \bibinfo
  {author} {\bibfnamefont {H.}~\bibnamefont {Kawano-Furukawa}}, \bibinfo
  {author} {\bibfnamefont {T.}~\bibnamefont {Saito}}, \bibinfo {author}
  {\bibfnamefont {A.}~\bibnamefont {Iyo}}, \bibinfo {author} {\bibfnamefont
  {H.}~\bibnamefont {Eisaki}}, \bibinfo {author} {\bibfnamefont
  {H.}~\bibnamefont {Fukazawa}}, \bibinfo {author} {\bibfnamefont
  {Y.}~\bibnamefont {Kohori}}, \bibinfo {author} {\bibfnamefont
  {K.}~\bibnamefont {Suzuki}}, \bibinfo {author} {\bibfnamefont
  {H.}~\bibnamefont {Usui}}, \bibinfo {author} {\bibfnamefont {K.}~\bibnamefont
  {Kuroki}}, \ and\ \bibinfo {author} {\bibfnamefont {K.}~\bibnamefont
  {Yamada}},\ }\href {\doibase 10.1103/PhysRevLett.106.067003} {\bibfield
  {journal} {\bibinfo  {journal} {Phys. Rev. Lett.}\ }\textbf {\bibinfo
  {volume} {106}},\ \bibinfo {pages} {067003} (\bibinfo {year}
  {2011})}\BibitemShut {NoStop}%
\bibitem [{\citenamefont {Zhang}\ \emph {et~al.}(2010)\citenamefont {Zhang},
  \citenamefont {Ma}, \citenamefont {Hou}, \citenamefont {Zhang}, \citenamefont
  {Xia}, \citenamefont {Chen}, \citenamefont {Hu}, \citenamefont {Luke},\ and\
  \citenamefont {Yu}}]{Zhang10PRB}%
  \BibitemOpen
  \bibfield  {author} {\bibinfo {author} {\bibfnamefont {S.~W.}\ \bibnamefont
  {Zhang}}, \bibinfo {author} {\bibfnamefont {L.}~\bibnamefont {Ma}}, \bibinfo
  {author} {\bibfnamefont {Y.~D.}\ \bibnamefont {Hou}}, \bibinfo {author}
  {\bibfnamefont {J.}~\bibnamefont {Zhang}}, \bibinfo {author} {\bibfnamefont
  {T.-L.}\ \bibnamefont {Xia}}, \bibinfo {author} {\bibfnamefont {G.~F.}\
  \bibnamefont {Chen}}, \bibinfo {author} {\bibfnamefont {J.~P.}\ \bibnamefont
  {Hu}}, \bibinfo {author} {\bibfnamefont {G.~M.}\ \bibnamefont {Luke}}, \ and\
  \bibinfo {author} {\bibfnamefont {W.}~\bibnamefont {Yu}},\ }\href {\doibase
  10.1103/PhysRevB.81.012503} {\bibfield  {journal} {\bibinfo  {journal} {Phys.
  Rev. B}\ }\textbf {\bibinfo {volume} {81}},\ \bibinfo {pages} {012503}
  (\bibinfo {year} {2010})}\BibitemShut {NoStop}%
\end{thebibliography}

\end{document}